\documentclass[journal,UTF8]{IEEEtran}

\usepackage{mathrsfs}
\usepackage[noadjust]{cite}
\usepackage{graphicx,color,overpic,psfrag}
\usepackage{float}
\usepackage{stfloats}
\usepackage{amsmath, amssymb}
\usepackage{cleveref}
\usepackage{latexsym}
\usepackage{bm}
\usepackage{amssymb}
\usepackage{cases}
\usepackage{array}
\usepackage{fancyhdr}
\usepackage{setspace}

\ifCLASSOPTIONcompsoc
\usepackage[caption=false,font=normalsize,labelfont=sf,textfont=sf]{subfig}
\else
\usepackage[caption=false,font=footnotesize]{subfig}
\fi

\definecolor{myblue}{rgb}{.91,.95,.99}
\usepackage{multirow}
\usepackage{url}
\usepackage{algorithm,algpseudocode,float}

\usepackage{blkarray}
\usepackage{booktabs}

\usepackage{multirow}
\usepackage{dsfont}
\usepackage{tabularx}
\usepackage[table]{xcolor}
\usepackage{amsfonts}

\usepackage{letltxmacro}

\usepackage{pifont}

\graphicspath{{figure/}}

\algdef{SE}[DOWHILE]{Do}{doWhile}{\algorithmicdo}[1]{\algorithmicwhile\ #1}
\graphicspath{{figure/}}

\newcommand{\ra}[1]{\renewcommand{\arraystretch}{#1}}
\newcolumntype{L}{>{\hspace*{-\tabcolsep}}l}
\newcolumntype{R}{c<{\hspace*{-\tabcolsep}}}
\definecolor{lightblue}{rgb}{0.93,0.95,1.0}

\newtheorem{prop}{Proposition}

\newcommand{\figref}[1]{Fig. \ref{#1}}
\newcommand{\tabref}[1]{Table \ref{#1}}
\newcommand{\alref}[1]{Algorithm \ref{#1}}

\newcommand{\secref}[1]{Section \ref{#1}}
\newcommand{\subsecref}[1]{Subsection \ref{#1}}

\newcommand{\propref}[1]{Proposition \ref{#1}}

\newcommand{\Exp}{{\mathsf{E}}}
\newcommand{\expect}[1]{\Exp\left\{#1\right\}}

\newcommand{\tr}[1]{\mathsf{tr}\left\{#1\right\}}
\newcommand{\diag}[1]{\mathsf{diag}\left\{#1\right\}}

\newcommand{\thetabs}[2]{{\dnnot{\theta}{bs}}}

\newcommand{\Lda}{\mathbf{\Lambda}}

\newcommand{\cC}{\mathcal{C}}
\newcommand{\cD}{\mathcal{D}}

\newcommand{\cN}{\mathcal{N}}
\newcommand{\cO}{\mathcal{O}}
\newcommand{\cP}{\mathcal{P}}
\newcommand{\cQ}{\mathcal{Q}}

\newcommand{\cU}{\mathcal{U}}

\newcommand{\cX}{\mathcal{X}}

\newcommand{\bn}{\mathbf{n}}

\newcommand{\bu}{\mathbf{u}}
\newcommand{\bv}{\mathbf{v}}

\newcommand{\bx}{\mathbf{x}}
\newcommand{\by}{\mathbf{y}}
\newcommand{\bz}{\mathbf{z}}

\newcommand{\bA}{\mathbf{A}}
\newcommand{\bB}{\mathbf{B}}
\newcommand{\bC}{\mathbf{C}}
\newcommand{\bD}{\mathbf{D}}
\newcommand{\bE}{\mathbf{E}}

\newcommand{\bG}{\mathbf{G}}
\newcommand{\bH}{\mathbf{H}}
\newcommand{\bI}{\mathbf{I}}

\newcommand{\bM}{\mathbf{M}}

\newcommand{\bP}{\mathbf{P}}
\newcommand{\bQ}{\mathbf{Q}}
\newcommand{\bR}{\mathbf{R}}

\newcommand{\bU}{\mathbf{U}}
\newcommand{\bV}{\mathbf{V}}
\newcommand{\bW}{\mathbf{W}}

\newcommand{\C}{\mathbb{C}}
\newcommand{\R}{\mathbb{R}}

\newcommand{\bzero}{\mathbf{0}}

\newcommand{\dnnot}[2]{#1_{\mathrm{#2}}}

\newcommand{\ntb}{\notag\\}

\newcommand{\Omegau}{\mathbf{\Omega}_{u}}

\allowdisplaybreaks

\UseRawInputEncoding

\begin{document}

\title{Energy Efficiency Maximization of Massive MIMO Communications With Dynamic Metasurface Antennas }
\author{

Li~You, Jie~Xu, George~C.~Alexandropoulos, Jue~Wang, Wenjin~Wang, and~Xiqi~Gao

\thanks{Copyright (c) 2015 IEEE. Personal use of this material is permitted. However, permission to use this material for any other purposes must be obtained from the IEEE by sending a request to pubs-permissions@ieee.org.}

\thanks{Part of this work was presented at the IEEE Global Communications Conference (GLOBECOM) 2021 \cite{2021DMAsEE}.}%

\thanks{
Li~You, Jie~Xu, Wenjin~Wang, and Xiqi~Gao are with the National Mobile Communications Research Laboratory, Southeast University, Nanjing 210096, China, and also with the Purple
Mountain Laboratories, Nanjing 211100, China (e-mail: lyou@seu.edu.cn; xujie@seu.edu.cn; wangwj@seu.edu.cn; xqgao@seu.edu.cn).
}
\thanks{
George~C.~Alexandropoulos is with the Department of Informatics and Telecommunications, National and Kapodistrian University of Athens, 15784 Athens, Greece and also with the Technology Innovation Institute, 9639 Masdar City, Abu Dhabi, United Arab Emirates (e-mail: alexandg@di.uoa.gr).
}
\thanks{
Jue~Wang is with School of Information Science and Technology, Nantong University, Nantong 226019, China, and also with Nantong Research Institute for Advanced Communication Technologies, Nantong 226019, China (e-mail:  wangjue@ntu.edu.cn).
}
}
\maketitle

\begin{abstract}
Future wireless communications are largely inclined to deploy massive numbers of antennas at the base stations (BSs) by leveraging cost- and energy-efficient as well as environmentally friendly antenna arrays. The emerging technology of dynamic metasurface antennas (DMAs) is promising to realize such massive antenna arrays with reduced physical size, hardware cost, and power consumption. The goal of this paper is the optimization of the energy efficiency (EE) performance of DMA-assisted massive multiple-input multiple-output (MIMO) wireless communications. Focusing on the uplink, we propose an algorithmic framework for designing the transmit precoding of each multi-antenna user and the DMA tuning strategy at the BS to maximize the EE performance, considering the
availability of either instantaneous or statistical channel state information (CSI). Specifically, the proposed framework is shaped around Dinkelbach's transform, alternating optimization, and deterministic equivalent methods. In addition, we obtain a closed-form solution to the optimal transmit signal directions for the statistical CSI case, which simplifies the corresponding transmission design for the multiple-antenna case. Our numerical results verify the good convergence behavior of the proposed algorithms, and showcase the considerable EE performance gains of the DMA-assisted massive MIMO transmissions over the baseline schemes.
\end{abstract}


\section{Introduction}\label{sec:Instruction}

Future wireless communications are expected to satisfy very high requirements, such as ultra-low latencies, high spectral efficiency (SE), and ultra-large connection, thus presenting a series of new challenges in the 5th generation (5G) mobile communication technology and beyond era \cite{2017Key}. Massive multiple-input multiple-output (MIMO) is a promising method to support such requirements by setting a massive number of antennas at the base station (BS), which has been proven to significantly increase the throughput of wireless systems \cite{2015Massive}. However, it brings a great demand on the radio frequency (RF) chains to realize massive MIMO transmissions by conventional antennas with fully digital architectures, which exposes some problems that cannot be ignored in practice, such as increased fabrication cost \cite{2016PhaseorSwitches}, high power consumption \cite{2017Hybrid}, limited physical size and shape, and deployment restriction \cite{Hoeher2017A}. To this end, some works have focused on the design of antennas to implement effective massive MIMO systems. Recent years have witnessed the increasing interest in an emerging antenna technology named dynamic metasurface antennas (DMAs), which is promising to realize practical massive antenna arrays for future wireless communications \cite{2021Dynamic6G}.

DMA is a brand-new concept for aperture antenna designs that leverage a kind of resonant, sub-wavelength, and tunable metamaterial elements to generate the desired radiations \cite{2018A}, \cite{2019Smart}. Specifically, each metamaterial element acts as a magnetic or electric polarizable dipole. When they are clustered in a planar surface, 
their collection can often be characterized by an effective permeability and permittivity. By introducing simplified tailored inclusions, the physical properties of each metamaterial, especially the permittivity and permeability, can be reconfigured to show a series of desired characteristics. Based on this feature, the planar structures can carry out different abilities of controllable signal processing, including radiation, amplified reflection, beamforming, and reception \cite{2012Holloway,2013Metamaterial,2006Metamaterials,2017Analysis}. Utilizing their reflection functionality, the planar structures termed as reconfigurable intelligent surfaces can overcome non-line-of-sight conditions of the propagation environments and improve the communication coverage effectively and energy-efficiently \cite{2020RISChannelEstimation,2021RISYuan,2021Reconfigurable,2020Multicell,2019ReconfigurableHuang,2020Holographicsurface,2021RISrich,2021Hybrid,2021ARISGC,Calvanese2021Recoonfigurable,2022JianRIS,2021StrWireless}. Moreover, when realizing radiation, beamforming, and receiving of signals, the planar structures are combined with waveguides generating a new paradigm for antennas that we focus on in this paper.

To appreciate the practical values of DMAs, we delve into their features and advantages over some existing technologies. As is mentioned above, future BSs tend to accommodate a massive number of antennas. However, conventional fully-digital transceivers connect each of the antenna elements to an individual RF chain. When such a transceiver with a massive number of antennas is used in future BSs, the size, power consumption, and hardware cost of the transceivers will be largely increased  \cite{2015Sidelobe}. By contrast, the number of RF chains required in DMA-based transceivers is much smaller than that in conventional transceivers, typically equal to the number of waveguides. Therefore, the physical area and power consumption of DMA-based transceivers can be significantly reduced, which makes it appealing for future green communications. Meanwhile, the independent data streams processed by a DMA-based transceiver are much fewer than metamaterial elements in the digital domain, which means that DMA-based transceivers enable a form of hybrid analog/digital (A/D) precoding. Compared with conventional hybrid A/D beamforming architectures that require numerous phase shifters to connect the antenna elements and RF chains, DMA-based hybrid A/D precoding does not require any additional analog combining circuitries. Specifically, the tuning of
metamaterial elements is often accomplished
with simple components, such as varactors, thus resulting in increased flexibility and reduced power consumption in the DMA-based hybrid A/D precoding \cite{2021Dynamic6G}.

Since DMAs can realize low-cost, power-efficient, and compact planar arrays, many studies have been conducted on their applications to implement massive MIMO systems in recent years. For example, authors in \cite{2019Enhancing} studied DMA-assisted spatial multiplexing wireless communications and demonstrated that DMAs could significantly enhance the capacity in MIMO channels with one or two clusters. Authors in \cite{9054184} studied the application of DMAs for MIMO orthogonal frequency division modulation (OFDM) receivers with bit-limited analog-to-digital converters (ADCs). The results showed that the DMA-based receivers with bit-limited ADCs were capable of recovering the transmit OFDM signals. Authors in \cite{8815427} and \cite{2019Dynamicuplink} respectively investigated the DMA tuning strategies for the uplink and downlink massive MIMO systems. Although DMAs are promising for MIMO communications, most of the existing works focused on DMA-based SE optimization, while DMA-based energy efficiency (EE) optimization has rarely been explored.

It is worth noting that most of the aforementioned works assumed that the instantaneous channel state information (CSI) is perfectly known for transmission design. DMA weight parameters are designed to adapt to the available channel states to improve the communication quality. Thus, with the perfectly known instantaneous CSI, the DMA-assisted systems can achieve a high capacity gain. However, tuning DMAs via exploiting instantaneous CSI 
is inappropriate and inadvisable due to the following reasons. Firstly, instantaneous CSI can be fast time-varying, which forces DMAs to frequently adjust their properties to keep up with the channel states, thus resulting in significant signaling overhead \cite{2021OverheadRIS}. Secondly, DMAs are equipped with smart controllers for realizing amplitude or phase tuning \cite{2020TowardsRIS}. Although the smart controllers operate under a tiny amount of energy, they are still power-consuming when overloaded with continuous operations, i.e., frequent tuning would not be energy efficient for DMAs. Therefore, when channels are fast time-varying, it is more reasonable and feasible to exploit the statistical CSI in DMA-assisted systems, which varies over larger time scales and results in less power consumption compared to exploiting instantaneous CSI.

Motivated by the above concerns, in this paper, we study the energy-efficient transmit precoding and DMA tuning strategies for a single-cell multi-user DMA-assisted massive MIMO uplink system. It is noted that in our previous work \cite{2021DMAsEE} we only studied the case with instantaneous CSI availability. In this paper, we make more substantial contributions, which are summarized as follows:
\begin{itemize}
  \item We study the EE maximization of the single-cell multi-user DMA-assisted massive MIMO uplink communications with instantaneous and statistical CSI, respectively. For both cases, we develop a well-structured and low-complexity algorithm framework for the transmit precoding design and DMA tuning strategy, including the deterministic equivalent (DE), Dinkelbach's transform, and the alternating optimization (AO) methods.
  \item For the case where instantaneous CSI is perfectly known, we develop an AO-based optimization framework to alternatingly update the transmit covariance matrices\footnote{Note that optimizing the transmit covariance matrix is a canonical way in the multi-user MIMO communications \cite{2003Introduction}. Actually, the transmit precoding matrix is embedded in the transmit covariance matrix under the context of eigenmode transmission.} of the multi-antenna users and the DMA weight matrix at the BS. For the transmit covariance design, we apply Dinkelbach's transform to solve the concave-linear fractional problem.
      For the DMA weights design, we firstly obtain the weight matrix in a closed form by neglecting its physical structure, and then adopt an AO-based algorithm to reconfigure it.
  \item To tackle the bottleneck of obtaining instantaneous CSI, we exploit statistical CSI to design the transmission strategy. Firstly, we derive an optimal closed-form solution to the transmit signal directions of users. Then, we apply the DE method to asymptotically approximate the ergodic SE, aiming to reduce the computational overhead. Next, we adopt Dinkelbach's transform to obtain the users' power allocation matrices. Finally, we derive the weight matrix of DMAs with a similar method to the instantaneous CSI case.
  \item Our extensive numerical results showcase the computational efficiency of our proposed EE optimization framework over benchmark schemes. It can be concluded that DMA-assisted massive MIMO communications can achieve higher EE performance than those based on conventional antennas, especially in the high power budget region.
\end{itemize}

The rest of the paper is organized as follows: \secref{sec:system model} illustrates the DMA input-output relationship and the channel model. \secref{sec:EE Optimization with Instantaneous CSIT} and \secref{sec:EE Optimization with Statistical CSI} investigate the considered EE maximization problem of the DMA-assisted MIMO uplink communications with instantaneous and statistical CSI, respectively. \secref{sec:numerical_result} provides our simulation results. Finally, \secref{sec:conclusion} concludes this paper.

The notations used throughout the paper are defined as follows: Boldface lower-case letters denote column vectors, e.g., $\bx$, and boldface upper-case letters denote matrices, e.g., $\bM$. The notation $\bzero$ denotes a zero vector or matrix, and $\bM \succeq \bzero$ denotes a positive semi-definite matrix. The notations $\cX $, $\C$, and $\R$ denote sets, sets of complex numbers, and sets of real numbers, respectively. The superscripts $(\cdot)^{-1}$, $(\cdot)^H$, $(\cdot)^T$, and $(\cdot)^*$ represent the matrix inverse, conjugate-transpose, transpose, and conjugate, respectively. The operators $\tr{\cdot}$, $\expect{\cdot}$, $\diag{\cdot}$, and $|\bM|$ represent matrix trace, expectation, diagonalization, and determinant of matrix $\bM$, respectively. The operator $\operatorname{Re}\left( \cdot \right)$ means the real part of the input, and the operator $||\cdot||_{\text{F}}$ means the Frobenius norm of the input. The operator $\odot$ denotes Hadamard product. The notations $\jmath$ and $\cO$ denote the imaginary unit and computational complexity, respectively.

\section{System Model}\label{sec:system model}

Our work considers a single-cell massive MIMO uplink system where the BS simultaneously receives signals from multiple users. In the following, we illustrate the input-output relationship of DMAs and the channel model.
\begin{figure}
	\centering
    \includegraphics[width=0.48\textwidth]{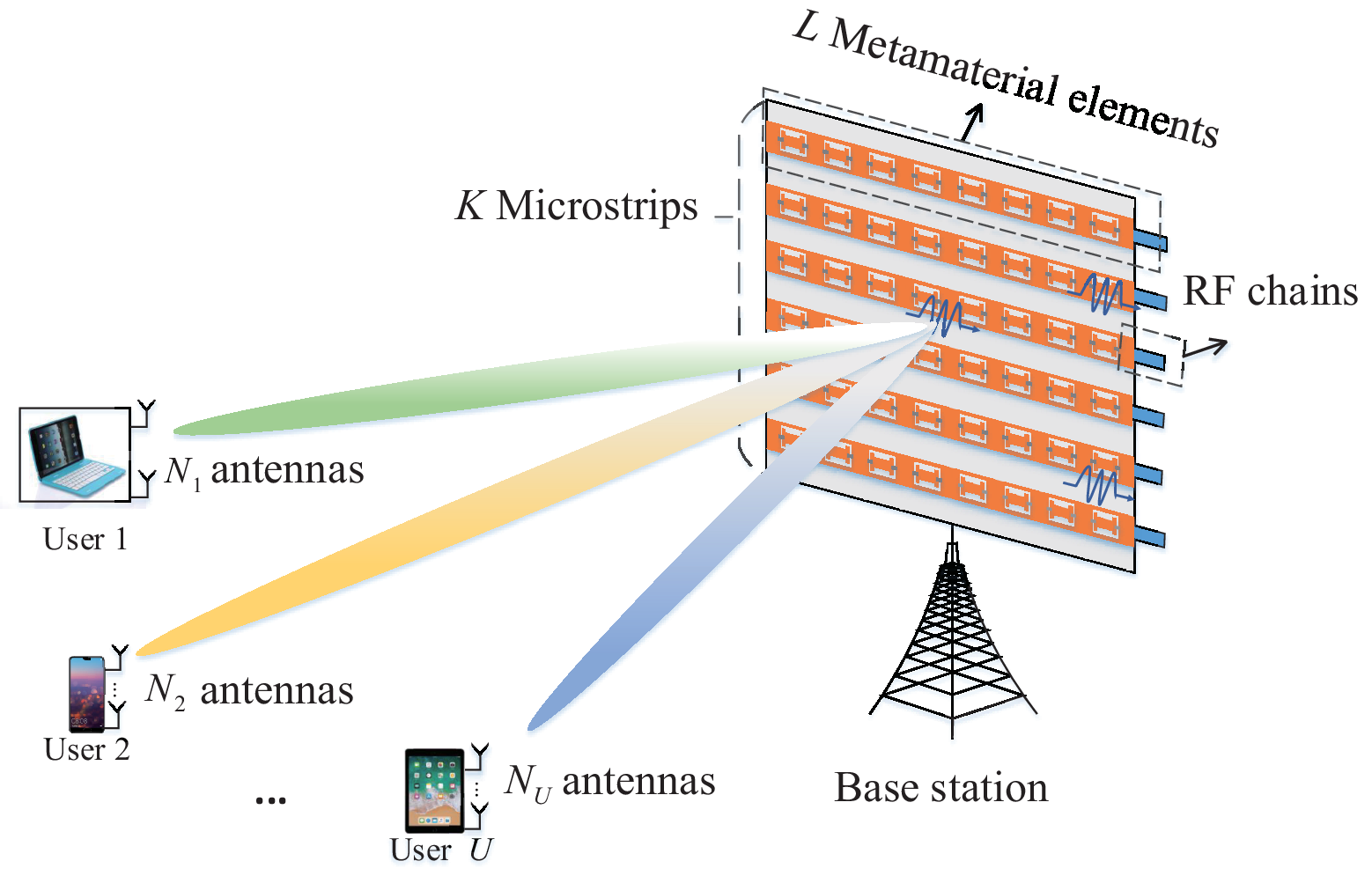}
	\caption{The considered DMA-assisted massive MIMO uplink system.}
	\label{fig:system_model}
\end{figure}
\subsection{Dynamic Metasurface Antennas}\label{subsec:Dynamic Metesuface Antennas}
As is shown in \figref{fig:system_model}, the considered system is composed of a DMA-based BS and $U$ users. The BS is equipped with a planar array consisting of $M$ metamaterial elements, and each user has an uniform linear array comprising $N_u$ conventional antennas interconnected via a fully digital beamforming architecture.
We define $\cU \triangleq  \{1,2,\ldots,U\}$ as the user set and $N_u$ as the number of conventional antennas at user $u \in \cU$. We assume that the DMA array consists of $K$ microstrips, e.g., the guiding structure whose top layer is embedded with metamaterials, and each microstrip consists of $L$ metamaterial elements, i.e., $M = K L$. Each metamaterial element observes the radiations from the channel, adjusts, and transmits them along the microstrip to the corresponding RF chain independently. The output signal of each microstrip is the linear combination of all the radiation observed by the corresponding $L$ metamaterial elements \cite{2019Dynamicuplink}.

We denote $\by\in {\C ^{M \times 1}}$ as the DMA input signals where $(\by)_{(k-1)L + l}$, $k \in \{1, \ldots ,K\}$, $l \in \{1, \ldots,L\}$ represents the observed radiation of the $l$th metamaterial in the $k$th microstrip. According to \cite{2021Dynamic6G}, the metamaterial element acts as a resonant electrical circuit and can be modeled as a causal filter. We use $h_{k,l}\in \C$ to denote the filter coefficient of the $l$th metamaterial in the $k$th microstrip, and denote by $\bH\in {\C ^{M \times M}}$ a diagonal matrix where $\left(\bH\right)_{\left( k-1 \right)L+l, \left( k-1 \right)L+l} = h_{k,l}$. Besides, the configurable weight matrix of DMAs is denoted as  $\bQ\in {\C ^{K \times M}}$. Then, the output signals of DMAs can be formulated as \cite{2019Dynamicuplink}
    \begin{align}\label{eq:DMA_ model}
    \bz = \bQ \bH \by \in \C^{K \times 1}.
    \end{align}
In \eqref{eq:DMA_ model}, the DMA weight matrix $\bQ$ is formulated as
    \begin{align}\label{eq:Q_constraint}
    \left( \bQ \right)_{k_1,(k_2-1)L+l}=
        \left\{
            \begin{aligned}
                & q_{k_1,l},\quad k_1=k_2 \\
                & 0,\quad \quad \ k_1\ne k_2 \\
             \end{aligned}
        \right. ,
    \end{align}
where $ k_1$, $k_2 \in {\{1,2,\ldots,K\}}$, $l \in {\{1,2,\ldots,L\}}$, and $q_{k_1,l}$ is the gain of the $l$th metamaterial in the $k_1$th microstrip. Eq. \eqref{eq:Q_constraint} considers the fact that DMA arrays can be formed by tiling together a set of microstrips \cite{2019Dynamicuplink}. Hence, Eq. \eqref{eq:Q_constraint} is referred to as the physical structure constraint of DMAs. Actually, by slightly modifying \eqref{eq:Q_constraint}, the input-output relationship of any two-dimensional DMAs can be denoted by \eqref{eq:DMA_ model}.




\subsection{Channel Model}

We define $\bx_u \in {\C^{N_u \times 1}}$ as the transmit signals from user $u$ with zero mean and the transmit covariance matrix $\expect{\bx_u \bx_u^H} = \bP_u \in \C^{N_u \times N_u}$. Additionally, $\bx_u$ satisfies $\expect{\bx_u \bx^H_{u'}} = \bzero$, $\forall u \neq u'$, which represents that the input signals from different users are independent of each other. Then, the channel output signal $\by$ is given by
    \begin{align}\label{eq:channel_inout}
    \by = \sum_{u=1}^U{\bG_u \bx_u} + \bn \in {\C ^{M \times 1}}.
    \end{align}
In \eqref{eq:channel_inout}, $\bG_u \in \C^{M \times N_u}$ denotes the channel between user $u$ and the BS, and $\bn \in {\C^{M\times 1}}$ denotes the independently and identically distributed  (i.i.d.) noise with covariance $\sigma^2 \bI_M$, where $\sigma^2$ denotes the noise power and $\bI_M$ is an $M \times M$ identity matrix.

Note that mutual coupling between the metamaterial elements is ignored in \eqref{eq:channel_inout} for simplicity. For the general case incorporating the mutual coupling effect, the model in \eqref{eq:channel_inout} can be slightly modified as $\by = \bC\sum_{u=1}^U{\bG_u \bx_u} + \bn$ where $\bC$ is the coupling matrix \cite{2017Stefanmutual}. Then, the proposed approaches in subsequent sections can still be applied via treating $\bC\bG_u$ as the equivalent channel matrix of user $u$.

We adopt the jointly-correlated Rayleigh fading channel model, in which the correlation properties at the users and the BS are modeled jointly \cite{2009Statistical}. Then, the channel matrices $\bG_u $, $\forall u \in \cU$, can be formulated as
   \begin{align}\label{eq:beam_H}
    \bG_u = \bU_u \widetilde{\bG}_u \bV_u^H,\quad \forall u \in \cU.
    \end{align}
In \eqref{eq:beam_H}, $\bV_u \in {\C ^{N_u\times N_u}}$ and $\bU_u \in {\C ^{M \times M}}$ are both deterministic unitary matrices, representing the eigenvectors of the transmit and receive correlation matrices, respectively \cite{2009Statistical}. In addition, $\widetilde{\bG}_u \in {\C^{M \times N_u}}$ represents the beam domain channel matrix, whose entries are zero-mean and independently Gaussian distributed. The channel statistics of $\widetilde{\bG}_u$ can be modeled as
    \begin{align}\label{eq:CSI}
    \bm{\Omega}_u = \expect{ \widetilde{\bG}_u \odot \widetilde{\bG}_u^* } \in { \R ^{M \times N_u}} .
    \end{align}
In \eqref{eq:CSI}, the entry, $[\bm{\Omega}_u]_{m,n}$, denotes the average energy coupled by the $m$th column entries of $\bU_u$ and the $n$th column entries of $\bV_u$. Hence, $\mathbf{\Omega}_u$ is also named as the eigenmode channel coupling matrix \cite{2009Statistical}.

Since DMAs act as receive antennas at the BS in our considered uplink, they observe and process signals from channels, i.e., channel output signals are fed directly to DMAs. With the input-output relationship of DMAs and the channel model given by \eqref{eq:DMA_ model} and \eqref{eq:channel_inout}, respectively, the relationship between the channel input and the DMA output can be given by
    \begin{align}
    \bz = \sum_{u=1}^{U}{\tilde{\bH}_u \bx_u + \tilde{\bn}} \in \C^{K \times 1},
    \end{align}
where $\tilde{\bH}_u \triangleq {\bQ \bH \bG_u }$ and $\tilde{\bn} \triangleq \bQ \bH \bn$.

\section{ EE Optimization With Instantaneous CSI}\label{sec:EE Optimization with Instantaneous CSIT}

In this section, we study the EE optimization of our DMA-assisted MIMO uplink system via exploiting the instantaneous CSI.\footnote{Note that the instantaneous CSI in DMA-based wireless communications can be obtained with the aid of some existing channel estimation methods for hybrid A/D wireless communications \cite{2018CSIHybrid,2019CSIHybrid}.} We firstly introduce the EE definition of our considered system. Then, we focus on designing the transmit covariance matrices $\bP_u$, $\forall u \in \cU$, and the DMA weight matrix $\bQ$ to maximize the system EE performance.

\subsection{Problem Formulation}
To define the system EE, we start with the SE definition of the DMA-assisted uplink system. Assume that all metamaterial elements have the same frequency selectivity, then $\bH$ can be expressed as $\bI$ multiplied by a constant \cite{2019Dynamicuplink}. Therefore, the achievable system SE is given by \cite{2021Reconfigurable,2019Dynamicuplink}
   \begin{align}\label{eq:sum rate_instantaneous}
    R =  \log_2 \left| \bI_K + \frac{1}{\sigma^2}\sum_{u=1}^U {\bQ \bG_u \bP_u \bG_u^H \bQ^H (\bQ \bQ^H)^{-1} } \right|.
    \end{align}

The whole power consumption of the DMA-assisted system consists of three major parts, including the transmit power, static hardware power, and dynamic power. Referring to \cite{2021Reconfigurable,2019Enhancing}, the whole power consumption of the DMA-assisted system is given by
\begin{align}\label{eq:power_consumption_model}
    W = \sum_{u=1}^{U}{( \xi_u \tr {\bP_u} + W_{\text{c},u})} + W_{\text{BS}} + K W_{\text{S}}.
\end{align}
In $\eqref{eq:power_consumption_model}$, $\xi_u=\rho_u^{-1}$ where $\rho_u$ denotes the transmit power amplifier efficiency of user $u$. $\tr{\bP_u}$ and $W_{\text{c},u}$ denote the transmit power consumption and static circuit power dissipation of user $u$, respectively. $W_{\text{S}}$ represents the dynamic power dissipation of each RF chain (e.g., power consumption in the ADCs, amplifier, and mixer). $W_{\text{BS}}$ incorporates the static circuit power dissipation at the BS. Note that the number of RF chains in the conventional antenna array with a fully digital transceiver architecture is equal to that of antenna elements. However, the number of RF chains in the DMA-assisted architecture is only equal to that of microstrips, resulting in the reduced dynamic power consumption by a factor of $L$ \cite{2019Enhancing}. In addition, the conventional antenna array with a hybrid A/D architecture also allows a reduced demand on RF chains. However, additional power consumption is required to support the phase shifters or switches.

With the system SE in \eqref{eq:sum rate_instantaneous} and power consumption in \eqref{eq:power_consumption_model}, the EE of our considered DMA-assisted uplink system is defined as
\begin{align}\label{eq:EE_model}
  EE =B \frac{ R}{W},
\end{align}
where $B$ is a constant denoting the channel bandwidth.
So far, the EE maximization problem of the DMA-assisted uplink system by designing the transmit covariance matrices $\bP_u$, $\forall u$, and DMA weight matrix $\bQ$ is formulated as follows:
\begin{subequations}\label{eq:instantaneous_model_P1}
     \begin{align}
       \cP_1: & \ntb
       \underset{\bQ,\bP}{ \mathop{\max }} \quad & \frac{ \log_2 \left| \bI_K + \frac{1}{\sigma^2}\sum\limits_{u=1}^U {\bQ \bG_u \bP_u \bG_u^H \bQ^H (\bQ \bQ^H)^{-1} } \right| }{\sum\limits_{u=1}^{U}{( \xi_u \tr {\bP_u} + W_{\text{c},u})} + W_{\text{BS}} + K W_{\text{S}}}, \label{eq:instantaneous_model_P1_a}\\
       \mathrm{s.t.} \quad & \left( \mathbf{Q} \right)_{k_1,(k_2-1)L+l} =
           \left\{
            \begin{aligned}
                & q_{k_1,l},\quad k_1=k_2 \\
                & 0,\quad \quad \  k_1\ne k_2 \\
             \end{aligned}
            \right. , \label{eq:instantaneous_model_P1_b}\\
          & \tr{ \bP_u } \leq P_{\max},\quad \bP_u \succeq \bzero, \quad \forall u \in \cU,\label{eq:instantaneous_model_P1_c}
      \end{align}
      \end{subequations}
where $P_{\max}$ denotes the maximum available transmit power. In addition, $\bP \triangleq \{\bP_1, \bP_2, \ldots, \bP_U\}$, $k_1$, $k_2 \in \{1, 2, \ldots ,K\}$, $l \in \{1, 2, \ldots, L\}$. In \eqref{eq:instantaneous_model_P1_a}, we ignore the constant $B$ without loss of generality. Problem $\cP_1$ is challenging to tackle with due to the following reasons. Firstly, since the objective function in \eqref{eq:instantaneous_model_P1_a} exhibits a fractional form, $\cP_1$ is an NP-hard problem \cite{2015EnergyZappone}. Secondly, the structure constraint of $\bQ$ in \eqref{eq:instantaneous_model_P1_b} is non-convex, which further complicates $\cP_1$. Thirdly, since variables $\bP$ and $\bQ$ are nonlinearly coupled, it is complicated to design $\bP$ and $\bQ$ simultaneously. To simplify the optimization process, we adopt an AO method to design $\bP$ and $\bQ$ in an alternating manner. For the optimization of $\bP$, we adopt Dinkelbach's transform to convert the concave-linear fraction in \eqref{eq:instantaneous_model_P1_a} into a concave one. For the optimization of $\bQ$, we first neglect constraint \eqref{eq:instantaneous_model_P1_b} to obtain the corresponding unconstrained $\bQ$, and then adopt an alternating minimization algorithm to reconfigure $\bQ$ to be constrained by \eqref{eq:instantaneous_model_P1_b}. Note that when $\xi_u$, $\forall u \in \cU,$ in \eqref{eq:instantaneous_model_P1_a} is equal to zero, the denominator of the objective function
is converted to a constant, and $\cP_1$ is reduced into a SE optimization problem. Thus, problem $\cP_1$ can describe both the EE and SE maximization problems of the considered DMA-assisted uplink communications.

\subsection{Optimization of the Unconstrained Weight Matrix}\label{subsec:Optimization of Unconstrained Weight Matrix}
When optimizing $\bQ$ with an arbitrarily given $\bP$, the denominator of $\eqref{eq:instantaneous_model_P1_a}$ can be treated as a constant. Thus, we only focus on the numerator maximization of $\eqref{eq:instantaneous_model_P1_a}$, i.e., the SE maximization.
By defining $\bar{\bG} = \frac{1}{\sigma^2}\sum_{u=1}^U{ \bG_u \bP_u \bG_u^H }$, and applying Sylvester's determinant identity $\log_2 \left| \bI+\bA\bB \right|= \log_2 \left| \bI+\bB\bA \right|$, the numerator of \eqref{eq:instantaneous_model_P1_a} can be written as
     \begin{align}\label{eq:rate_2_perf}
         R = \log_2 \left| \bI_M + \bar{\bG} \bQ^H \left( \bQ \bQ^H \right)^{-1} \bQ  \right|.
      \end{align}
Let $\bar{\bV}_{1}$ denote the right singular vectors matrix of $\bQ$, and $\bar{\bV}_{2}$ denote the first $K$ columns of $\bar{\bV}_{1}$. According to the projection matrix property that $ \bQ^H \left( \bQ \bQ^H \right)^{-1} \bQ  = \bar{\bV}_2 \bar{\bV}_{2}^H$ \cite{2000Matrix}, Eq. \eqref{eq:rate_2_perf} can be written as
     \begin{align}\label{eq:rate_projection}
         R = \log_2 \left| \bI_K + \bar{\bV}_{2}^H \bar{\bG} \bar{\bV}_{2} \right|.
     \end{align}

With the non-convex constraint in \eqref{eq:instantaneous_model_P1_b}, Eq. \eqref{eq:rate_projection} is difficult to tackle directly. Hence, we drop constraint \eqref{eq:instantaneous_model_P1_b} and consider a relaxed version of problem $\cP_1$. Then, when designing $\bQ$ with a given $\bP$, problem $\cP_1$ is recast as follows
     \begin{align}
       \cP_2:\quad \underset{\bar{\bV}_2}{ \mathop{\max }} \quad \log_2 \left| \bI_K + \bar{\bV}_{2}^H \bar{\bG} \bar{\bV}_{2}\right|. \label{problem_model_2_a}
     \end{align}
The solution to $\cP_2$ can be obtained in a close form according to \emph{\propref{theorem:RS_OD}}, as follows.
\begin{prop}\label{theorem:RS_OD}
Let $\bar{\bV}_3$ denote the eigenvectors corresponding to the largest $K$ eigenvalues of $\bar{\bG}$. Then, the maximal achievable SE in \eqref{problem_model_2_a} can be achieved by setting $\bar{\bV}_2$ as $\bar{\bV}_3$, i.e.,
\begin{align}
\bar{\bV}_2 = \bar{\bV}_3.
\end{align}
\end{prop}
The proof of \emph{\propref{theorem:RS_OD}} is similar to \cite[Corollary 2]{2019Dynamicuplink}, thus is omitted here.

By the singular value decomposition (SVD), the DMA weight matrix $\bQ$ can be written as
  \begin{align}\label{eq:Q_Close_1}
         \bQ  = \bar{\bU}_{2} \bar{\bD}_{2} \bar{\bV}_{2}^H,
     \end{align}
where $\bar{\bU}_{2} \in \C^{K \times K}$ and $\bar{\bD}_{2} \in \C^{K \times K}$ denote the left singular vector matrix and the diagonal singular value matrix of $\bQ$, respectively. From \emph{\propref{theorem:RS_OD}}, we can find that the maximal SE in \eqref{eq:rate_2_perf} only depends on the right singular vector matrix $\bar{\bV}_2$ and is independent of $\bar{\bU}_{2}$ and $\bar{\bD}_{2}$. Thus, we can design $\bar{\bU}_{2}$ and $\bar{\bD}_{2}$ to obtain $\bQ$ constrained by \eqref{eq:instantaneous_model_P1_b}.

\subsection{Optimization of the Transmit Covariance Matrices}\label{subsec:Optimization of Transmit Covariance Matrices}
When designing the transmit covariance matrices $\bP_u, \forall u \in \cU,$ with a given $\bar{\bV}_2$, problem $\cP_1$ is recast as
     \begin{subequations}\label{problem_model_only_P_1}
     \begin{align}
         \cP_3: \quad
         \underset{\bP}{\mathop{\max }}&\quad \frac{\log_2 \left| \bI_K + \frac{1}{\sigma^2}\sum\limits_{u=1}^{U}{\bar{\bV}_{2}^H \bG_u \bP_u \bG_u^H \bar{\bV}_{2}}\right|}
       { \sum\limits_{u=1}^{U}{( \xi_u \tr {\bP_u} + W_{\text{c},u})} + W_{\text{BS}} + M W_{\text{S}}}, \label{problem_model_only_P_a_1}\\
          \mathrm{s.t.}& \quad \tr{ \bP_u } \leq P_{\max},\quad \bP_u \succeq \bzero,\quad \forall u \in \cU .
      \end{align}
      \end{subequations}
Eq. \eqref{problem_model_only_P_a_1} is a concave-linear fraction whose numerator is concave and denominator is linear with respect to $\bP_u, \forall u$. Dinkelbach's transform is a classical method to address this kind of problems, and is guaranteed to converge to the optimal solution to $\cP_3$ with a super-linear rate \cite{2015EnergyZappone}. By invoking Dinkelbach's transform, problem $\cP_3$ is transformed to
   \begin{subequations} \label{eq:P3}
    \begin{align}
     \cP_4: \quad \underset{\bm{\bP},\eta_1 } {\mathop{ \max}} \quad & R(\bP) -\eta_1 W{(\bP)}, \\
     \quad \mathrm{s.t.}\quad &\tr{\bm{\bP}_u} \leq P_{\max},\; \bm{\bP}_u \succeq \bzero,\; \forall u \in \cU.
    \end{align}
  \end{subequations}
In $\cP_4$, $ R(\bP)$ and $W{(\bP)}$ respectively denote the numerator and denominator of \eqref{problem_model_only_P_a_1}, and $\eta_1$ is an auxiliary variable.  Problem $\cP_4$ can be addressed by alternatingly optimizing $\bP$ and $\eta_1$. With an arbitrarily given $\eta_1$, the optimal $\bP$ can be obtained by classical convex optimization techniques \cite{2004ConvexBoyd}.
Meanwhile, the optimal $\eta_1$ with an arbitrarily given $\bP$ is obtained by
    \begin{equation}\label{eq:eta_1}
         \eta_1^{*}=\frac{ R(\bP)}{W(\bP) }.
    \end{equation}
More details about this procedure based on Dinkelbach's transform are summarized in \textbf{\alref{alg:Dinkelbach's Transform}}.
    \begin{algorithm}[t]
    \caption{Dinkelbach's Transform} 
     \label{alg:Dinkelbach's Transform}
     \begin{algorithmic}[1]
    \Require The right singular vector matrix $\bar{\bV}_2$, threshold $\epsilon$.
    \State Initialize $\eta_1^{(\ell)}$ and set iteration index $\ell=0$.  
     \Repeat
    　　\State Set $\ell = \ell+1$.
        \State  Calculate $\bm{\bP}^{(\ell)}$ in \eqref{eq:P3} with $\eta_1^{(\ell-1)}$.
        \State Calculate $\eta_1^{(\ell)}$ in \eqref{eq:eta_1} with $\bm{\bP}^{(\ell)}$.
     \Until{$\left| \eta_1^{(\ell)}- \eta_1^{(\ell-1)} \right|\leq \epsilon $}
    \Ensure The optimal transmit covariance matrices $\bm{\bP} =\bm{\bP} ^{(\ell)}. $
    \end{algorithmic}
    \end{algorithm}

\subsection{Optimization of the Constrained Weight Matrix}\label{subsec:Optimization of Constrained Weight Matrix}

As is illustrated in \subsecref{subsec:Optimization of Unconstrained Weight Matrix}, the maximal SE of $\cP_2$ is independent of the unitary matrix $\bar{\bU}_{2}$ and diagonal matrix $\bar{\bD}_{2}$. Referring to \cite{2019Dynamicuplink}, we adopt an alternating minimization algorithm to adjust $\bar{\bU}_2$, $\bar{\bD}_{2}$, and $\bQ$. Let $\cQ_2^{K \times M}$ denote the set of $K \times M$ matrices conforming to \eqref{eq:instantaneous_model_P1_b}, $\cU^K$ denote the set of $K \times K$ unitary matrices, and $\cD^K$ denote the set of $K \times K$ diagonal matrices with positive diagonal entries.
The corresponding alternating approximation problem is given by
   \begin{align}\label{eq:Q_alternating}
     \cP_5: \quad
    &\underset{\bQ \in \cQ_2^{K \times M}, \bar{\bU}_{2} \in \cU^K, \bar{\bD}_2 \in \cD^K }{\mathop{\min }}\ \left\|  \bQ - \bar{\bU}_{2} \bar{\bD}_{2} \bar{\bV}_{3}^H \right\|^2_{\text{F}}.
    \end{align}
The detailed calculation of $\bQ$, $\bar{\bU}_{2}$, and $\bar{\bD}_2$ are described as follows.
    \begin{algorithm}[t]
    \caption{Alternating Minimization for the DMA Weights} 
     \label{alg:Alternating_Algorithm_for_DMA_Weights}
     \begin{algorithmic}[1]
    \Require The right singular vector matrix $\bar{\bV}_3$, threshold $\epsilon$.
    \State Initialize the iteration index $\ell=0$, $\bar{\bU}_2^{(\ell)}=\bI_K$ and $\bar{\bD}_2^{(\ell)}=\bI_K$.    
    \Repeat 
        \State Set $\ell = \ell+1$.
        \State Set $\bQ^{(\ell)}= \bQ^{\text{AM}}$ with $\bM = \bar{\bU}_2^{(\ell-1)} \bar{\bD}_2^{(\ell-1)} \bar{\bV}_3^H $ using \eqref{eq:PQ_AM}.
        \State Set $\bar{\bU}_2^{(\ell)}= \bar{\bU}_2^{\text{AM}}$ with $\bM_1 = \bQ^{(\ell)}$ and $\bM_2 = \bar{\bD}_2^{(\ell-1)} \bar{\bV}_3^H$ using \eqref{eq:U_AM}.
        \State Set $\bar{\bD}_2^{(\ell)}= \bar{\bD}_2^{\text{AM}}$ with $\bM_1 = (\bar{\bU}_2^{(\ell)})^H  \bQ^{(\ell)}$ and $\bM_2 = \bar{\bV}_3^H$ using \eqref{eq:D_AM}.
    \Until{$\left\| \bQ^{(\ell)}- \bQ^{(\ell-1)} \right\|_{\text{F}} \leq \epsilon $}
    \Ensure The weight matrix $ \bQ = \bQ^{(\ell)} $.
    \end{algorithmic}
    \end{algorithm}

Firstly, we define $\bM \triangleq \bar{\bU}_{2} \bar{\bD}_{2} \bar{\bV}_{3}^H$. With arbitrarily given $\bar{\bU}_{2}$ and $\bar{\bD}_{2}$, we can obtain $\bQ$ by solving
   \begin{subequations}
    \begin{align}\label{eq:Q_AM}
    {{\mathbf{Q}}^{\text{AM}}}\left( \mathbf{M} \right)\triangleq \underset{\mathbf{Q}\in \cQ_2^{K \times M}}{\mathop{\arg \min }}\ {{\left\|\bQ  -\bM \right\|}^{2}_{\text{F}}}.
    \end{align}
By defining $\cQ$ as the set of possible values for the entries of $\bQ$, we have
    \begin{align}\label{eq:PQ_AM}
& \left({\bQ^{\text{AM}}(\bM)} \right)_{k_1,(k_2-1)L+l} \ntb
      & =  \left\{
            \begin{aligned}
                & \underset{q\in \cQ }{\mathop{\arg \min }}\, {\left| q-{{\left( \mathbf{M} \right)}_{k_1, \left( {{k}_{2}}-1 \right)L+l}} \right|^{2}}, \quad k_1=k_2 \\
                & 0,\quad \quad \quad \; \quad \quad \quad \quad \quad \quad\quad\quad\quad\quad k_1\ne k_2 \\
             \end{aligned}
        \right. .
    \end{align}

Secondly, we define $\bM_1 = \bQ $ and $\bM_2 = \bar{\bD}_2 \bar{\bV}_3^H$.  By letting $\tilde{\bU}$ and $\tilde{\bV}$ be the left and right singular vector matrices of $\bM_1 \bM_2^H$, respectively, we can obtain $\bar{\bU}_{2}$ with arbitrarily given $\bar{\bD}_{2}$ and $\bQ$ via
    \begin{align}\label{eq:U_AM}
    \bar{\bU}_{2}^{\text{AM}} \left( {\mathbf{M}}_1,{\mathbf{M}}_2 \right) & \triangleq \underset{\bar{\bU}_{2} \in \cU^K }{\mathop{\arg \min }}\ {{\left\| {\mathbf{M}}_1-\bar{\bU}_{2}{\mathbf{M}}_2 \right\|}^{2}_\text{F}}  =\tilde{\bU}\tilde{\bV}^{H}.
    \end{align}

Finally, we define $\bM_1 = \bar{\bU}_2^H  \bQ$ and $\bM_2 = \bar{\bV}_3^H$. By letting $\mathbf{m}_{1,i}$ and $\mathbf{m}_{2,i}$ denote the $i$th columns of $\bM_1^H$ and $\bM_2^H$, respectively, we can obtain $\bar{\bD}_{2}$ with arbitrarily given $\bar{\bU}_{2}$ and $\bQ$ via
    \begin{align}\label{eq:D_AM_1}
    \bar{\bD}_{2}^{\text{AM}} \left( {{\mathbf{M}}_1},{{\mathbf{M}}_2} \right)\triangleq \underset{\bD_{2} \in \cD^K }{\mathop{\arg \min }}\ {{\left\| {{\mathbf{M}}_1}-\bar{\bD}_2{{\mathbf{M}}_2} \right\|}^{2}_\text{F}}.
    \end{align}
In \eqref{eq:D_AM_1}, the diagonal entries of $\bar{\bD}_{2}^{\text{AM}}$ are given by
    \begin{align}\label{eq:D_AM}
    {\left( \bar{\bD}_{2}^{\text{AM}} \left( {\mathbf{M}}_1,{\mathbf{M}_2} \right) \right)}_{i,i}=\max \left( \frac{\operatorname{Re}\left( \mathbf{m}_{1,i}^H {\mathbf{m}_{2,i}} \right)}{{{\left\| {\mathbf{m}_{2,i}} \right\|}^{2}_\text{F}}},\delta  \right),
    \end{align}
   \end{subequations}
where $\delta$ is a small positive number \cite{2019Dynamicuplink}.

Problem $\cP_5$ can be addressed by alternatingly calculating \eqref{eq:PQ_AM}, \eqref{eq:U_AM}, and \eqref{eq:D_AM}. The alternating minimization algorithm for DMA weight design is summarized in \textbf{\alref{alg:Alternating_Algorithm_for_DMA_Weights}}.

\subsection{Convergence and Complexity Analysis}
So far, we have studied the EE maximization problem $\cP_1$ of the DMA-assisted MIMO uplink communications with instantaneous CSI. The approaches for designing users' transmit covariance matrices and the DMA weight matrix are described in \subsecref{subsec:Optimization of Unconstrained Weight Matrix}, \subsecref{subsec:Optimization of Transmit Covariance Matrices} and \subsecref{subsec:Optimization of Constrained Weight Matrix}, respectively. Now, we present the complete AO-based algorithm to find the transmit covariance matrices and the DMA weight matrix in \textbf{\alref{alg:overall_instantaneous}}.
\begin{algorithm}[t]
    \caption{AO-based Algorithm for EE Maximization With Instantaneous CSI} 
     \label{alg:overall_instantaneous}
     \begin{algorithmic}[1]

    \Require The channel matrices $\bG_u, \forall u$, the noise power $\sigma^2$, power consumptions $W_{\text{c},u}$, $W_{\text{BS}}$, and $W_{\text{S}}$, threshold $\epsilon$.

    \State Initialize the iteration index $\ell=0$, the unconstrained right singular matrix $\bar{\bV}_2^{(\ell)}$, and the EE performance $EE^{(\ell)}$.
    \Repeat
        \State $\ell= \ell+1 $. 
        \State Obtain $\bP_u^{(\ell)}, \forall u,$ with $\bar{\bV}_2^{(\ell-1)}$ and \textbf{\alref{alg:Dinkelbach's Transform}}.
        \State Obtain $\bar{\bV}_2^{(\ell)}$ with $\bP_u^{(\ell)}, \forall u,$ and \emph{\propref{theorem:RS_OD}}.
        \State Update $EE^{(\ell)}$ using $\bP_u^{(\ell)}, \forall u,$ and $\bar{\bV}_2^{(\ell)}$.
    \Until{$\left| EE^{(\ell)}- EE^{(\ell-1)} \right| \leq \epsilon $}
    \State Obtain the weight matrix $\bQ$ with     $\bar{\bV}_2^{(\ell)}$ and \textbf{\alref{alg:Alternating_Algorithm_for_DMA_Weights}}.
    \Ensure The weight matrix $\bQ$ and the transmit covariance matrices $\bP_u = \bP_u^{(\ell)}, \forall u$.
    \end{algorithmic}
    \end{algorithm}

In \textbf{\alref{alg:overall_instantaneous}}, $\bQ$ and $\bP_u$, $\forall u$, are alternatingly optimized. In particular, $\bP_u$, $\forall u$, is obtained by Dinkelbach's method, which is guaranteed to converge to the global optimum of the fractional program in $\cP_3$ \cite{2015EnergyZappone}.
In addition, $\bQ$, $\bar{\bU}_2$, and $\bar{\bD}_2$ can be iteratively obtained in close forms, as shown in \eqref{eq:PQ_AM}, \eqref{eq:U_AM}, and \eqref{eq:D_AM},
As the Frobenius norm objective in (19) is differentiable, the convergence of the alternating optimization for $\bQ$ is guaranteed \cite{2019Dynamicuplink}. Hence, the proposed AO-based Algorithm for EE Maximization with instantaneous CSI in \textbf{\alref{alg:overall_instantaneous}} is guaranteed to converge.

The main structure of \textbf{\alref{alg:overall_instantaneous}} includes an AO method for alternatingly designing $\bP$ in $\cP_4$ and $\bar{\bV}_2$ in $\cP_2$ and \textbf{\alref{alg:Alternating_Algorithm_for_DMA_Weights}} for alternatingly designing $\bQ$, $\bar{\bU}_2$ and $\bar{\bD}_2$
in $\cP_5$.
Firstly, we discuss the complexity of the AO-based algorithm for optimizing $\bP$ and unconstrained $\bar{\bV}_2$. For the transmit covariance matrices $\bP$ optimized by Dinkelbach's method, we assume that the optimization process requires $I_{\text{DK1}}$ iterations. Since each iteration needs to optimize $\sum_{u=1}^U{N_u^2}$ variables and the complexity per iteration is polynomial over the number of variables \cite{2001Lectures}, the complexity of optimizing the transmit covariance matrices $\bP$ is estimated as $\cO(I_{\text{DK1}}(\sum_{u=1}^U{N_u^2})^p)$, where $1 \leq p \leq 4$ \cite{2019ReconfigurableHuang}. For optimizing $\bar{\bV}_2$ in problem $\cP_2$, it requires only one iteration. The computational complexity mainly depends on the eigenvalue decomposition of $\frac{1}{\sigma^2}\bG_u \bP_u \bG_u^H \in\C ^{M \times M}$. Thus, the complexity of optimizing $\bar{\bV}_2$ is estimated as $\cO(M^3)$, which is small and negligible compared with that of optimizing $\bP$. Therefore, the complexity of the AO-based algorithm for optimizing $\bP$ and unconstrained $\bar{\bV}_2$ is estimated as $\cO(I_{\text{AO}}I_{\text{DK1}}(\sum_{u=1}^U{N_u^2})^p)$, where $I_{\text{AO}}$ is number of required iterations in the AO method.
Then, for \textbf{\alref{alg:Alternating_Algorithm_for_DMA_Weights}}, the computational complexity per iteration depends on the complexity of calculating $\bQ$, $\bar{\bU}_2$, and $\bar{\bD}_2$ in \eqref{eq:PQ_AM}, \eqref{eq:U_AM}, and \eqref{eq:D_AM}, respectively, which is estimated as $\cO(M^3)$.
Hence, with the assumption that \textbf{\alref{alg:Alternating_Algorithm_for_DMA_Weights}} requires $I_{\text{C}}$ iterations, the complexity of \textbf{\alref{alg:Alternating_Algorithm_for_DMA_Weights}}
can be estimated as $\cO(I_{\text{C}} M^3)$.
Therefore, the computational complexity of the proposed EE maximization algorithm for the considered DMA-assisted MIMO uplink with instantaneous CSI is estimated as $\cO(I_{\text{AO}}I_{\text{DK1}}(\sum_{u=1}^U{N_u^2})^p + I_{\text{C}} M^3)$.

\section{ EE Optimization With Statistical CSI}\label{sec:EE Optimization with Statistical CSI}

Channels might be fast time-varying in practical wireless communications, thus frequently tuning DMAs and reallocating transmit power with instantaneous CSI might be difficult. In such cases, utilizing statistical CSI to optimize the system EE performance is more efficient \cite{2009Statistical}. In this section, we explore approaches to optimize the system EE by designing the transmit covariance matrices and DMA weight matrix via exploiting statistical CSI.

\subsection{Problem Formulation}
To formulate the corresponding EE maximization problem, we firstly describe the system SE and power consumption metrics.
For the statistical CSI case, we adopt the ergodic achievable SE metric defined as
  \begin{align}\label{eq:sum rate_statistical}
    \bar{R} =  \expect{\log_2 \left| \bI_K + \frac{1}{\sigma^2}\sum_{u=1}^U {\bQ \bG_u \bP_u \bG_u^H \bQ^H (\bQ \bQ^H)^{-1} } \right|},
    \end{align}
where the expectation is taken over the channel realizations. 

In addition, we use \eqref{eq:power_consumption_model} to model the overall power consumption.
Then, the corresponding EE maximization problem can be formulated as
\begin{subequations}\label{eq:statistical_model_P1}
     \begin{align}
     \bar{\cP}_1: & \ntb
         \underset{\bQ,\bP}{ \mathop{\max }}& \quad \frac{ \expect{\log_2 \left| \bI_K + \frac{1}{\sigma^2}\sum\limits_{u=1}^U {\bQ \bG_u \bP_u \bG_u^H \bQ^H (\bQ \bQ^H)^{-1} } \right|} }{\sum\limits_{u=1}^{U}{( \xi_u \tr {\bP_u} + W_{\text{c},u})} + W_{\text{BS}} + KW_{\text{S}}},\label{eq:statistical_model_P1_a}\\
       \mathrm{s.t.} & \quad \left( \mathbf{Q} \right)_{k_1,(k_2-1)L+l}=
           \left\{
            \begin{aligned}
                & q_{k_1,l},\quad k_1=k_2 \\
                & 0,\quad \quad \ k_1\ne k_2 \\
             \end{aligned}
            \right. , \label{eq:statistical_P1_b}\\
          & \quad \tr{ \bP_u } \leq P_{\max},\quad \bP_u \succeq \bzero, \quad \forall u \in \cU, \label{eq:statistical_P1_c}
      \end{align}
      \end{subequations}
where $k_1$, $k_2 \in \{1, 2, \ldots, K\}$ and $l \in \{1, 2, \ldots, L\}$. Note that in problem $\bar{\cP}_1$, we utilize the same power consumption notations $W_{\text{c},u}$, $W_{\text{BS}}$, and $W_{\text{S}}$ as those in the instantaneous CSI case. Since they are all constants, they will not affect the following optimization development. $\bar{\cP}_1$ is challenging to tackle because \eqref{eq:statistical_model_P1_a} exhibits a concave-linear fractional structure and \eqref{eq:statistical_P1_b} is a non-convex constraint. In addition, the expectation operation in \eqref{eq:statistical_model_P1_a} further increases the computational overhead. In the following, we aim to cope with the foregoing difficulties to obtain the EE maximization in $\bar{\cP}_1$. Note that when $\xi_u$, $\forall u \in \cU$, is set as zero, problem
$\bar{\cP}_1$ reduces to a SE optimization problem with statistical CSI.

\subsection{ Optimization of Users' Transmit Covariance Matrices}\label{subsec:Optimization_of_Users' Transmit_Covariance_Matrices_statistical}
In order to find $\bP$ which maximizes \eqref{eq:statistical_model_P1_a}, we apply the projection matrix property \cite{2000Matrix}. Then, the ergodic achievable SE in \eqref{eq:sum rate_statistical} can be reformulated as
     \begin{align}\label{eq:rate_projection_2}
         \bar{R} = \expect{\log_2 \left| \bI_K + \frac{1}{\sigma^2} \sum\limits_{u=1}^U{\bar{\bV}_{2}^H \bG_u \bP_u \bG_u^H \bar{\bV}_{2}} \right|},
     \end{align}
where $\bar{\bV}_2$ denotes the first $K$ columns of the right singular vector matrix of $\bQ$.
Similar to \secref{sec:EE Optimization with Instantaneous CSIT}, we adopt an AO method to optimize $\bP$ and $\bar{\bV}_{2}$ iteratively. We firstly consider the design of $\bP_u$, $\forall u \in \cU$, with an arbitrarily given $\bar{\bV}_{2}$. Then, problem $\bar{\cP}_{1}$ is recast as
     \begin{subequations}\label{problem_model_only_P}
     \begin{align}
        \bar{\cP}_{2}: \quad
\underset{\bP}{\mathop{\max }}& \quad \frac{\expect{ \log_2 \left| \bI_K + \frac{1}{\sigma^2}\sum\limits_{u=1}^U{\bar{\bV}_{2}^H \bG_u \bP_u \bG_u^H  \bar{\bV}_{2}} \right|}}
       { \sum\limits_{u=1}^{U}{( \xi_u \tr {\bP_u} + W_{\text{c},u})} + W_{\text{BS}} + K W_{\text{S}}}, \label{problem_model_only_P_a}\\
         \mathrm{s.t.} &\quad \ \tr{ \bP_u } \leq P_{\max},\quad \bP_u \succeq \bzero,\quad \forall u \in \cU .
      \end{align}
      \end{subequations}

Considering the high computational complexity of a large number of variables in $\bar{\cP}_{2}$, we decompose the transmit covariance matrices $\bP_u$, $\forall u \in \cU$ via eigenvalue decomposition, which is written as
     \begin{align}\label{eq:eigen}
         \bP_u = \bm{\Phi}_u \Lda_u \bm{\Phi}_u^H, \quad \forall u \in \cU.
     \end{align}
In \eqref{eq:eigen}, $\bm{\Phi}_u$ and $\Lda_u$ denote the transmit signal directions and the transmit power allocation of user $u$, respectively. We will respectively introduce the approaches for $\bm{\Phi}_u$ and $\Lda_u, \forall u \in \cU,$ in the following.

\subsubsection{Optimal Transmit Directions at Users}The optimal transmit signal directions can be obtained by the following proposition.
\begin{prop}\label{theorem:beam_domain_optimal}
The optimal transmit direction of user $u$ is identical to the eigenvector matrix of the transmit correlation matrix corresponding to the channel between user $u$ and the BS, i.e.,
    \begin{align}\label{eq:beam_optimal}
    \bm{\Phi}_u =  \bV_u.
    \end{align}
 \end{prop}
The proof of \emph{\propref{theorem:beam_domain_optimal}} is similar to \cite[Proposition 1]{2020EEDownlink}, thus is omitted here.

\emph{\propref{theorem:beam_domain_optimal}} indicates that the transmit precoding is aligned to the eigenvectors of the transmit correlation matrices to maximize the system EE. By applying \emph{\propref{theorem:beam_domain_optimal}}, the transmit covariance matrix of user $u$ is formulated as $\bP_u = \bV_u \Lda_u \bV_u^H$, $ \forall u \in \cU$. Then, problem $\bar{\cP}_2$ is formulated as
      \begin{subequations}\label{eq:problem_model_only_Lambda}
     \begin{align}
\bar{\cP}_{3}:  & \ntb
        \underset{\Lda}{\mathop{\max }}&\quad \frac{\expect{ \log_2 \left| \bI_K + \frac{1}{\sigma^2}\sum\limits_{u=1}^U{\bar{\bV}_{2}^H \bU_u \tilde{\bG}_u \Lda_u  \tilde{\bG}_u^H \bU_u^H \bar{\bV}_{2}} \right|}}
       { \sum\limits_{u=1}^{U}{( \xi_u \tr {\Lda_u} + W_{\text{c},u})} + W_{\text{BS}} + K W_{\text{S}}},  \label{eq:problem_model_only_Lambda_a}\\
            \mathrm{s.t.}&\quad \tr{\Lda_u} \leq P_{\max},\; \Lda_u \succeq \bzero,\; \Lda_u \ \text{diagonal}, \; \forall u \in \cU,
      \end{align}
      \end{subequations}
where $\Lda \triangleq \{\Lda_1, \Lda_2, \ldots, \Lda_U \}$. Since the transmit direction, $\bm{\Phi}_u, \forall u$, can be determined with a closed-form solution by \emph{\propref{theorem:beam_domain_optimal}},
the number of optimization variables has been significantly reduced.

\subsubsection{Deterministic Equivalent Method}
Problem $\bar{\cP}_3$ can be approximated by the Monte-Carlo method via averaging over a large number of samples, but this method is computationally expensive. Hence, we adopt the DE method, an asymptotic expression based on the large-dimensional random matrix theory, to approximate the expectation in \eqref{eq:problem_model_only_Lambda_a}. Notice that the adopted asymptotic approximation is sufficiently accurate for small-scale MIMO systems \cite{2011On}.

Define $\widehat{\bG}_u \triangleq \widehat{\bU}_u \widetilde{\bG}_u \widehat{\bV}_u^H \in \C^{K \times N_u}$, where $\widehat{\bU}_u \triangleq \bar{\bV}_2^H \bU_u \in \C^{K \times M}$, $\widehat{\bV}_u \triangleq \bI_{N_u}$, and $\widetilde{\bG}_u$ denotes the beam domain channel between user $u$ and the BS. Define $\widehat{\bG} \triangleq {[\widehat{\bG}_1,\widehat{\bG}_2,\ldots, \widehat{\bG}_U]} \in \C^{K \times N}$ and $\bD \triangleq \diag{\Lda_1, \Lda_2, \ldots \Lda_U}\in \C^{N \times N}$. Then, the numerator of $\eqref{eq:problem_model_only_Lambda_a}$ is written as
    \begin{align}\label{eq:sample_rate}
    \bar{R} =  \expect{ \log_2 \left| \bI_K + \frac{1}{\sigma^2}\widehat{\bG} \bD \widehat{\bG}^H \right|}.
    \end{align}

By adopting the DE method \cite{2011On}, Eq. \eqref{eq:sample_rate} can be approximated by
    \begin{align}\label{eq:DE_rate}
    R_{\text{DE}} = & \sum_{u=1}^U{\log_2 \left| \bI_{Nu}  + \bm{\Xi}_u \Lda_u \right|} + \log_2 \left| \bI_K + \bm{\Psi} \right| \ntb
     & - \frac{1}{\ln2} \sum_{u=1}^U {\bm{\gamma}_u^T \bm{\Omega}_u \bm{\psi}_u },
    \end{align}
where $\bm{\gamma}_u \triangleq [\gamma_{u,1}, \gamma_{u,2}, \ldots,\gamma_{u,M}]^T $, ${\bm{\psi}_u \triangleq [\psi_{u,1}, \psi_{u,2}, \ldots, }$ ${\psi_{u,N_u}]^T }$ and $\bm{\Psi} \triangleq \sum\limits_{u=1}^U{\bm{\Psi}_u} \in \C^{K \times K}$. The calculation of $\bm{\Xi}_u$ and $\bm{\Psi}_u$, $\forall u \in \cU$, are given by
    \begin{align}\label{eq:DE_rate_Xi_Psi}
    \left\{
     \begin{aligned}
        &\bm{\Xi}_u = \widehat{\bV}_u \diag{\bm{\Omega}_u^T \bm{\gamma}_u} \widehat{\bV}_u^H  \in \C^{N_u \times N_u}, \\
        &\bm{\Psi}_u = \frac{1}{\sigma^2}\widehat{\bU}_u \diag{\bm{\Omega}_u \bm{\psi}_u} \widehat{\bU}_u^H \in \C^{K \times K}.
      \end{aligned}
      \right.
    \end{align}
The quantities $\bm{\gamma} \triangleq \{\gamma_{u,m}\}_{\forall{u,m}}$ and $\bm{\psi} \triangleq \{\psi_{u,n}\}_{\forall{u,n}}$ form the unique solution to the equations
    \begin{align}\label{eq:DE_rate_gamma_phi}
    \left\{
     \begin{aligned}
        &\gamma_{u,m} = \frac{1}{\sigma^2}\widehat{\bu}_{u,m}^H (\bI_K + \bm{\Psi})^{-1} \widehat{\bu}_{u,m} , \\
        &\psi_{u,n} =\widehat{\bv}_{u,n}^H \Lda_u ( \bI_{Nu} + \bm{\Xi}_u \Lda_u)^{-1} \widehat{\bv}_{u,n},
     \end{aligned}
    \right.
    \end{align}
where $\widehat{\bv}_{u,m}$ is the $m$th column of $\widehat{\bV}_u$ and $\widehat{\bu}_{u,n}$ is the $n$th column of $\widehat{\bU}_u$. The detailed procedure of the DE method is presented in \textbf{\alref{alg:Deterministic_Equivalent_Method}}.

    \begin{algorithm}[t]
    \caption{Deterministic Equivalent Method}
    \label{alg:Deterministic_Equivalent_Method}
    \begin{algorithmic}[1]
    \Require The power allocation matrices $\Lda_u$, $\forall u \in \cU$, the right singular vector matrix $\bar{\bV}_2$ and threshold $\epsilon$.
    \For {{$u=1$} to {$U$}}
    \State Initialize the iteration index $\ell=0$ and $\bm{\psi}_u^{(\ell)}$.
    \Repeat
        \State Set $\ell = \ell+1$.
        \For {$m=1$ to $M$}
           \State Calculate $\gamma_{u,m}^{(\ell)}$ by \eqref{eq:DE_rate_gamma_phi} with $\bm{\psi}_u^{(\ell-1)}$.
        \EndFor
        \State Obtain: $\bm{\gamma}_u^{(\ell)} = [\gamma_{u,1}^{(\ell)}, \gamma_{u,2}^{(\ell)}, \ldots,\gamma_{u,M}^{(\ell)}]^T $.
        \For {$n=1$ to $N_u$}
           \State Calculate  $\psi_{u,N_u}^{(\ell)}$ by \eqref{eq:DE_rate_gamma_phi} with $\bm{\gamma}_u^{(\ell)}$.
        \EndFor
        \State Obtain: $\bm{\psi}_u^{(\ell)} = [\psi_{u,1}^{(\ell)}, \psi_{u,2}^{(\ell)}, \ldots,\psi_{u,N_u}^{(\ell)}]^T $.
     \Until{$\left\| \bm{\psi}_u^{(\ell)} -\bm{\psi}_u^{(\ell-1)} \right\|_{\text{F}} \le \epsilon $}
     \State Use $\bm{\psi}_u^{(\ell)}$ and $\bm{\gamma}_u^{(\ell)}$ to calculate $\bm{\Xi}_u$ and $\bm{\Psi}_u$ with \eqref{eq:DE_rate_Xi_Psi}.
    \EndFor
    \State Set $\bm{\gamma}_u = \bm{\gamma}_u^{(\ell)}$ and $\bm{\psi}_u = \bm{\psi}_u^{(\ell)}, \forall u \in \cU$, and use them to calculate $R_{\text{DE}}$ in \eqref{eq:DE_rate}.
     \Ensure The DE sum-rate $R_{\text{DE}}$ and auxiliary variables $\bm{\psi}_u$, $\bm{\gamma}_u$, $\bm{\Xi}_u$, $\bm{\Psi}, \forall u \in \cU $.
    \end{algorithmic}
    \end{algorithm}

By adopting the DE method, problem $\bar{\cP}_3$ is recast as
     \begin{subequations}\label{eq:problem_model_only_Lambda_with_DE}
     \begin{align}
         \bar{\cP}_{4}:\quad\underset{\Lda}{\mathop{\max}}\quad & \frac{R_{\text{DE}}(\Lda)}{W(\Lda)},\\
        \mathrm{s.t.}\quad & \tr{\Lda_u} \leq P_{\max},\; \Lda_u \succeq \bzero,\; \ntb
         &\Lda_u \ \text{diagonal}, \; \forall u \in \cU.
      \end{align}
      \end{subequations}
In $\bar{\cP}_4$, $R_{\text{DE}}(\Lda)$ and $W(\Lda)$ are functions of $\Lda$, denoting the asymptotic
SE in \eqref{eq:DE_rate} and the power consumption in the denominator of \eqref{eq:problem_model_only_Lambda_a}, respectively. Since variables $\Lda$ and  $\bm{\psi}$ are mutually related, $\bm{\psi}$ needs to be updated when $\Lda$ is updated.

\subsubsection{Transmit Power Allocation at Users}
Problem $\bar{\cP}_4$ is a classical concave-convex fractional program, so we invoke Dinkelbach's transform to convert it to a convex problem. Specifically, problem $\bar{\cP}_4$ is reformulated as
   \begin{subequations}\label{eq:Dinkelbach's_Lambda}
    \begin{align}
     \bar{\cP}_5: \quad  \underset{\Lda, \eta_2} {\mathop{\arg\max}} \quad  &R_{\text{DE}}(\Lda)-\eta_2 W(\Lda), \label{eq:Dinkelbach's_Lambda_a} \\
     \mathrm{s.t.}\quad & \tr{\Lda_u} \leq P_{\max},\; \Lda_u \succeq \bzero,\; \ntb
     & \Lda_u \ \text{diagonal}, \; \forall u \in \cU, \label{eq:Dinkelbach's_Lambda_b}
    \end{align}
  \end{subequations}
where $\eta_2$ is an auxiliary variable. Problem $\bar{\cP}_5$ can be efficiently tackled by optimizing $\Lda$ and $\eta_2$ in an alternating manner. When $\eta_2$ is given, the optimal $\Lda$ can be obtained by convex optimization techniques \cite{2004ConvexBoyd}. Meanwhile, with given $\Lda$, the optimal solution to $\eta_2$ is obtained by
    \begin{equation}\label{eq:Dinkelbach's_Lambda_eta}
         \eta_2^{*}=\frac{ R_{\text{DE}}(\Lda)}{W(\Lda)}.
    \end{equation}

The optimization process of $\bar{\cP}_5$ is similar to \textbf{\alref{alg:Dinkelbach's Transform}}. The main difference from \textbf{\alref{alg:Dinkelbach's Transform}} is that the optimization process of $\bar{\cP}_5$ adopts an asymptotic SE expression due to lacking the instantaneous CSI. In addition, in $\bar{\cP}_5$ we need to consider the interaction between $\Lda$ and $\bm{\psi}$, i.e., each time $\bm\Lambda$ is updated, $\bm{\psi}$ must be updated to ensure that the asymptotic SE in \eqref{eq:DE_rate} is valid.

\subsection{Optimization of the DMA Weight Matrix}\label{subsec:Optimization_of_DMAs_Weight_Matrix}
\subsubsection{Optimization of the Unconstrained Weight Matrix}
If the transmit covariance matrices are fixed, the denominator of the objective function in $\bar{\cP}_1$ is a constant. Hence, when optimizing  $\bQ$ with a given $\bP$, we only analyze the numerator of \eqref{eq:statistical_model_P1_a} and ignore the denominator for clarity. By applying the projection matrix property, DE method, and \emph{\propref{theorem:beam_domain_optimal}}, the numerator of \eqref{eq:statistical_model_P1_a} is approximated by \eqref{eq:DE_rate}. To maximize \eqref{eq:DE_rate} with a given $\Lda$, we optimize the variable $\bar{\bV}_2$ and the auxiliary variable $\bm{\psi}$ in an iterative manner. When optimizing $\bar{\bV}_2$ with given $\bm{\psi}$, only the second term of $ R_{\text{DE}}(\Lda)$, $\log_2 \left| \bI_K + \bm{\Psi} \right|$, is affected by $\bar{\bV}_2$, and the effect on the first and third terms of $ R_{\text{DE}}(\Lda)$ can be removed \cite{2011On}. Therefore, when $\bm{\psi}$ is given, we only consider the optimization of the second term, $\log_2 \left| \bI_K + \bm{\Psi} \right|$, with respect to $\bar{\bV}_2$, and the corresponding problem without constraint \eqref{eq:statistical_P1_b} is formulated as
     \begin{align}\label{eq:DE_rate_projection_1}
        \bar{\cP}_6: \quad  \underset{\bar{\bV}_2}{\mathop{\max}} \quad  {\log_2 \left| \bI_K + \frac{1}{\sigma^2}\sum_{u=1}^U{ \widehat{\bU}_u \diag{\Omegau \bm{\psi}_u} \widehat{\bU}_u^H } \right|},
     \end{align}
where $\widehat{\bU}_u = \bar{\bV}_2^H \bU_u$.

Define $\bA \triangleq  \frac{1}{\sigma^2}\sum_{u=1}^U{ \bU_u \diag{\Omegau \bm{\psi}_u} \bU_u^H}$, then the objective function in $\bar{\cP}_6$ is written as
     \begin{align}\label{eq:DE_rate_projection_2}
       R_{\text{DE},2} = \log_2 \left| \bI_K + \bar{\bV}_2^H \bA \bar{\bV}_2  \right|.
     \end{align}
Since \eqref{eq:DE_rate_projection_2} is identical with \eqref{eq:rate_projection}, a similar conclusion can be obtained from \emph{\propref{theorem:RS_OD}}, i.e., the
maximal $R_{\text{DE},2}$ can be obtained by setting $\bar{\bV}_2$ as the eigenvectors corresponding to the largest $K$ eigenvalues of $\bA$. By updating $\bar{\bV}_2$ and $\bm{\psi}$ alternatingly, we can obtain the optimal solution of $\bar{\cP}_6$.

\subsubsection{Optimization of the Constrained Weight Matrix}
By the SVD, the DMA weight matrix can be written as $\bQ =\bU_{2} \bD_{2} \bV_{2}^H$. Similarly to the instantaneous CSI case, we apply the alternating minimization algorithm to optimize $\bU_{2}$, $\bD_{2}$, and $\bQ$. The problem formulation and solution are the same as those in \subsecref{subsec:Optimization of Constrained Weight Matrix}, so we omit the detailed description. The alternating
minimization algorithm for optimizing $\bQ$ with constraint \eqref{eq:statistical_P1_b} can be found in \textbf{\alref{alg:Alternating_Algorithm_for_DMA_Weights}}.

\subsection{Convergence and Complexity Analysis }\label{subsec:overall_algorithm}
In the above two subsections, we have provided approaches for designing the transmit covariance matrices of the multi-antenna users and the DMA weight matrix at the BS with statistical CSI. Unlike \secref{sec:EE Optimization with Instantaneous CSIT}, we obtain the transmit directions of each user by a closed-form solution, which significantly reduces the number of variables. In addition, we apply the DE method to approximate the ergodic SE, thus further simplifying the optimization process. The overall algorithm to obtain the power allocation matrices of users and the DMA weight matrix is presented in \textbf{\alref{alg:Overall_algorithm_With_statitical_CSI}}.
    \begin{algorithm}[t]
    \caption{AO-based Algorithm for EE Maximization With Statistical CSI} 
     \label{alg:Overall_algorithm_With_statitical_CSI}
     \begin{algorithmic}[1]

    \Require Eigenmode channel coupling matrices $\mathbf{\Omega}_u$, $\forall u$, the noise power $\sigma^2$, power consumptions $W_{\text{c},u}$, $W_{\text{BS}}$, and $W_{\text{S}}$, threshold $\epsilon$.

    \State Initialize the iteration index $\ell_1=\ell_2=\ell_3=0$, the system EE $EE^{(\ell_3)}$, the transmit power allocation matrices $\Lda_u^{(\ell_1)}, \forall u$, the unconstrained right singular matrix $\bar{\bV}_2^{(\ell_2)}$.
    \Do
        \State $\ell_3=\ell_3+1 $.
        \State Obtain $\bm{\psi}^{(\ell_1+\ell_2)}$ via $\bV_2^{(\ell_2)}$, $\Lda_u^{(\ell_1)}, \forall u$, and \textbf{\alref{alg:Deterministic_Equivalent_Method}}.
        \State Obtain $\eta_2^{(\ell_1)}$ via $\bm{\psi}^{(\ell_1)},$ $\Lda_u^{(\ell_1)}, \forall u,$ and  \eqref{eq:Dinkelbach's_Lambda_eta}.
       \Repeat
            \State $\ell_1=\ell_1+1 $.
            \State Obtain $\Lda_u^{(\ell_1)}$ in $\bar{\cP}_5$ with $\bm{\psi}^{(\ell_1+\ell_2-1)}$ and $\eta_2^{(\ell_1-1)}$.
            \State Update $\bm{\psi}^{(\ell_1+\ell_2)}$ via $\Lda_u^{(\ell_1)}, \forall u,$ and \textbf{\alref{alg:Deterministic_Equivalent_Method}}.
            \State Update $\eta_2^{(\ell_1)}$ via $\bm{\psi}^{(\ell_1+\ell_2)},$ $\Lda_u^{(\ell_1)}, \forall u,$ and \eqref{eq:Dinkelbach's_Lambda_eta}.
        \Until{$\left| \eta_2^{(\ell_1)} - \eta_2^{\ell_1-1} \right| \leq \epsilon $}
        \Repeat
            \State  $\ell_2 =\ell_2 +1$
            \State Update $\bar{\bV}_2^{(\ell_2)}$ via $\bm{\psi}^{(\ell_1+\ell_2-1)}$ and \emph{\propref{theorem:RS_OD}}.
            \State Update $\bm{\psi}^{(\ell_1+\ell_2)}$ via $\bar{\bV}_2^{(\ell_2)}$  and \textbf{\alref{alg:Deterministic_Equivalent_Method}}.
        \Until{$\left\| \bar{\bV}_2^{(\ell_2)} - \bar{\bV}_2^{(\ell_2-1)} \right\|_{\text{F}} \leq \epsilon $}
        \State Update $EE^{(\ell_3)}$ via $\bar{\bV}_2^{(\ell_2)}$ and $\Lda_u^{(\ell_1)}, \forall u$.
    \doWhile{$\left| EE^{(\ell_3)}- EE^{(\ell_3-1)} \right| \geq \epsilon $}
    \State Set $\Lda_u = \Lda_u^{(\ell_1)}, \forall u$. Obtain the DMA weights $\bQ$ with $\bar{\bV}_{2}^{(\ell_2)}$ and \textbf{\alref{alg:Alternating_Algorithm_for_DMA_Weights}}.
    \Ensure The DMA weights $\bQ$, the transmit power allocation matrices $\Lda_u, \forall u$.
    \end{algorithmic}
    \end{algorithm}

In \textbf{\alref{alg:Overall_algorithm_With_statitical_CSI}},
$\bP_u$ is obtained by iteratively optimizing $\bm{\Phi}_u$ and $\bm{\Lambda}_u$, $\forall u$. Specifically, $\bm{\Phi}_u$, $\forall u$, is obtained in a close form by using \emph{Proposition 2}, and $\bm{\Lambda}_u$, $\forall u$, is optimized by Dinkelbach's transform. The result is guaranteed to converge to the optimum in $\bar{\cP}_2$ \cite{2015EnergyZappone}. In addition, $\bQ$ is obtained by solving the Frobenius norm objective in (19), whose convergence is guaranteed since the objective function is differentiable \cite{2019Dynamicuplink}. Therefore, the convergence of \textbf{\alref{alg:Overall_algorithm_With_statitical_CSI}} is guaranteed.

The complexity of \textbf{\alref{alg:Overall_algorithm_With_statitical_CSI}} depends on that of the AO-based method for alternatingly optimizing $\Lda$ and $\bar{\bV}_2$ and  \textbf{\alref{alg:Alternating_Algorithm_for_DMA_Weights}} for alternatingly optimizing $\bQ$, $\bar{\bU}_2$, and $\bar{\bD}_2$. For the AO-based algorithm, the per-iteration complexity mainly depends on optimizing $\Lda$ by Dinkelbach's transform. Meanwhile, the complexity of the DE method in \textbf{\alref{alg:Deterministic_Equivalent_Method}} and the closed-form calculation of $\bar{\bV}_2$ is very small, thus is ignored.
Assume that there are $I_{\text{DK2}}$ iterations for optimizing $\Lda$ by Dinkelbach's transform. Since the number of variables is $\sum_{u=1}^U{N_u}$ in each iteration, the complexity of optimizing $\Lda$ is estimated as $\cO(I_{\text{DK2}}(\sum_{u=1}^U{N_u})^p)$, where $1 \leq p \leq 4$ \cite{2019ReconfigurableHuang}. In addition, assume that the AO-based algorithm includes $I_{\text{AO}}$ iterations, then its complexity can be approximated by $\cO(I_{\text{AO}}I_{\text{DK2}}(\sum_{u=1}^U{N_u})^p)$. For alternatingly optimizing $\bQ$, $\bar{\bU}_2$, and $\bar{\bD}_2$ by \textbf{\alref{alg:Alternating_Algorithm_for_DMA_Weights}}, the complexity is estimated as $\cO(I_{\text{Q2}}M^3)$, where $I_{\text{Q2}}$ denotes the number of iterations number and $\cO(M^3)$ denotes the complexity per iteration. Hence, by exploiting the statistical CSI, the total complexity of the proposed EE maximization algorithm
is $\cO(I_{\text{AO}}I_{\text{DK2}}(\sum_{u=1}^U{N_u})^p+I_{\text{Q2}}M^3)$.


\section{Numerical Results}\label{sec:numerical_result}

This section provides numerical results to assess the proposed approach for the DMA-assisted multiuser MIMO uplink transmission. Our simulation adopts the QuaDRiGa normalization channel model, the 3GPP-UMa-NLoS propagation environment for small scale fading \cite{2014QuaDRiGa}, and assumes all the channels exhibit the same large scale fading factor as $-120$ dB \cite{2020EEDownlink}. The channel statistics, $\bm{\Omega}_u, \forall u,$ can be obtained by the existing methods, e.g., \cite{2009Statistical}. We set the number of users as $U = 6$ and each user is equipped with 4 antennas, i.e., $N_u = 4$, $\forall u \in \cU$. The antennas of users are placed in uniform linear arrays spaced with half wavelength. We set the number of microstrips as $K = 8$ and each microstrip is embedded with $L = 8$ metamaterial elements. The space between metamaterial elements on the DMA array is set as 0.2 wavelength. We set the bandwidth as $B = 10$ MHz, the amplifier inefficiency factor as $\rho = 0.3$, $\forall u$, and the noise variance as $\sigma^2 = -96$ dBm. For the power consumption, we set the static circuit power as $W_{\text{c},u} = 20$ dBm, $\forall u$, the hardware dissipated power at the BS as $W_{\text{BS}} = 40$ dBm, and the power consumption per RF chain as $W_{S} = 30$ dBm \cite{2021Reconfigurable}, \cite{2013Coordinated}. Additionally, the entries of the DMA weight matrix $\bQ$ can be selected from the following four sets \cite{2019Dynamicuplink}:
\begin{itemize}
  \item UC:$\ $ the complex plane, i.e., $\cQ= \C$;
  \item AO:$\ $ amplitude only, i.e., $\cQ= \left[0.001,5  \right]$;
  \item BA:$\ $ binary amplitude, i.e., $\cQ= \{0, 0.1\}$;
  \item LP:$\ $ Lorentzian-constrained phase, i.e., $\cQ= \{ \frac{\jmath+e^{\jmath \phi}}{2} \}$, $ \phi \in [0, 2\pi] $.
\end{itemize}
%
%
%
%

\subsection{Convergence Performance}
\begin{figure}
\centering
\subfloat[]{\centering\includegraphics[width=0.45\textwidth]{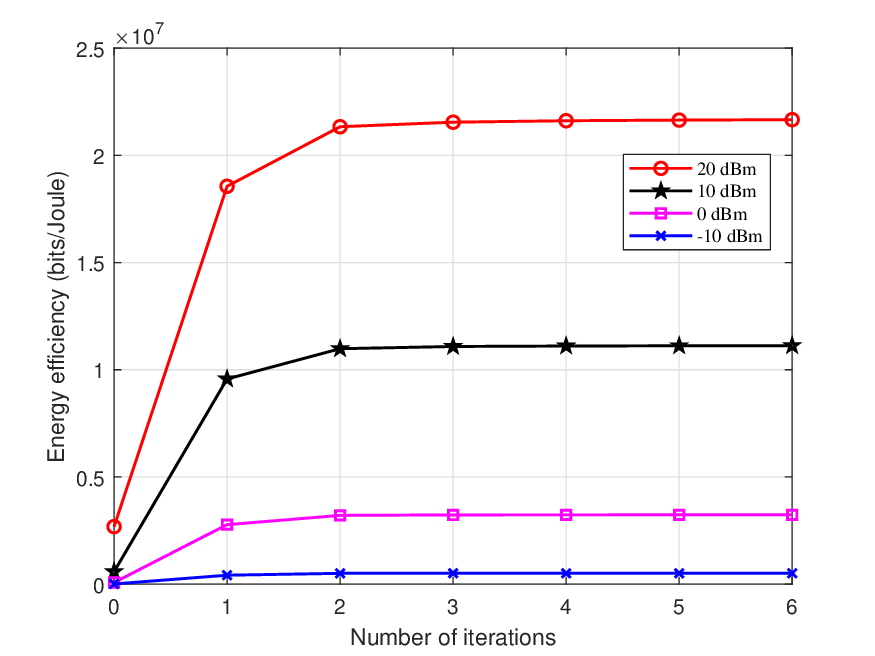}\label{fig:Convergence_per}}\hfill
\subfloat[]{\centering\includegraphics[width=0.45\textwidth]{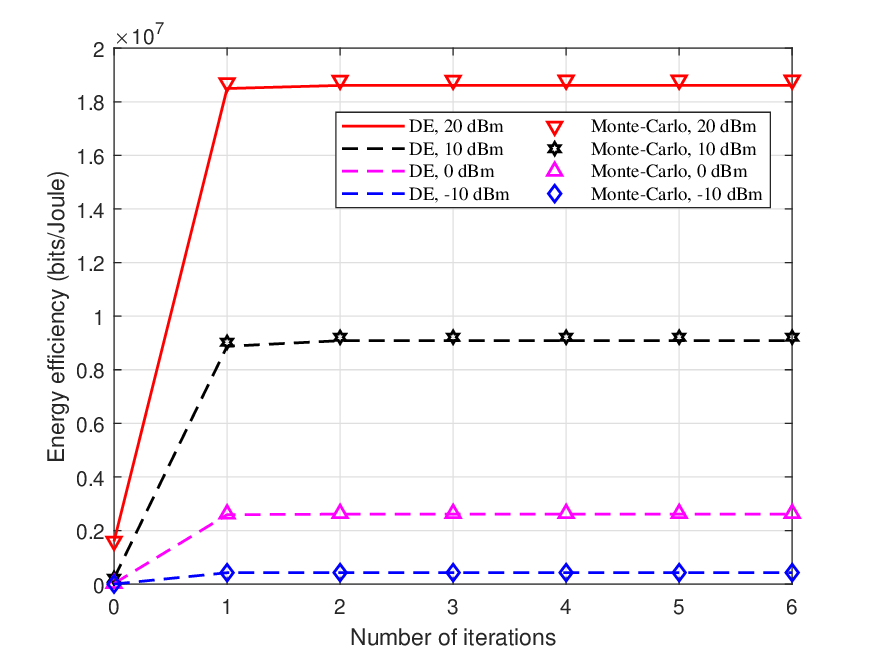}\label{fig:Convergence_sta}}
\caption{Convergence performance of the AO-based EE maximization algorithms with different power budgets: (a) instantaneous CSI; (b) statistical CSI.}
\label{fig:Convergence}
\end{figure}


The convergence performance of the proposed AO-based algorithms in the instantaneous and statistical CSI cases under different transmit power budgets are respectively presented in \figref{fig:Convergence}\subref{fig:Convergence_per} and \figref{fig:Convergence}\subref{fig:Convergence_sta}. For both cases, the proposed EE maximization algorithms converge at a rapid rate for different power budgets. Besides, \figref{fig:Convergence}\subref{fig:Convergence_sta} verifies the accuracy of the asymptotic SE expression. The gap between the DE-based and Monte-Carlo-based results is negligible. Thus, we confirm that adopting the DE method is valid and computationally efficient for resource allocation in the DMA-assisted MIMO communications with statistical CSI.

\subsection{EE Performance Comparison Between Instantaneous and Statistical CSI Cases}

\begin{figure}
\centering
\includegraphics[width=0.45\textwidth]{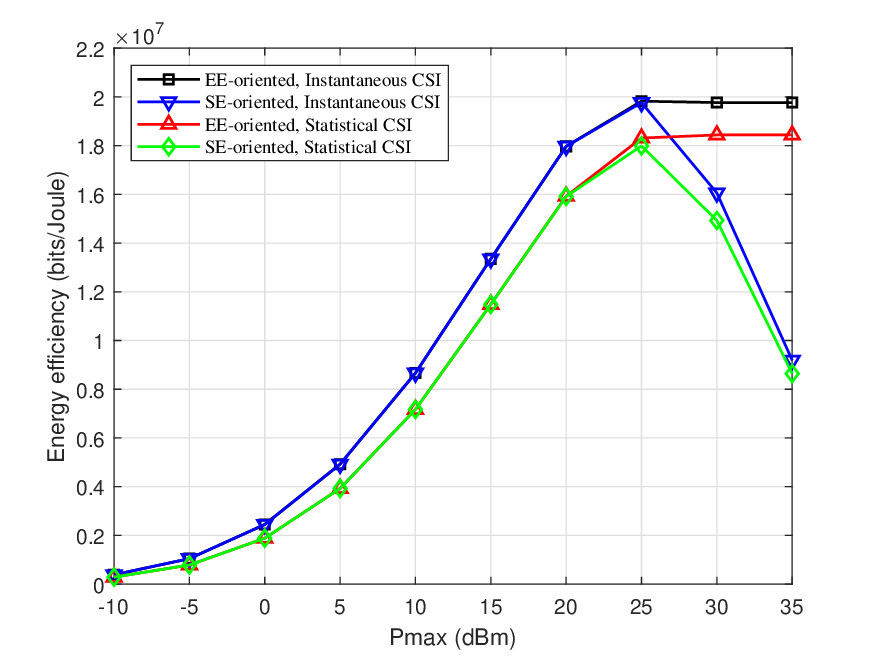}
\caption{EE performance comparison between the instantaneous and statistical CSI cases versus the transmit power budget in both the SE- and EE-oriented approaches.}
\label{fig:CSIComparison}
\end{figure}
In this subsection, we compare the EE performance of the DMA-assisted communications between the instantaneous and statistical CSI cases in the EE- and SE-oriented approaches, respectively. Note that ``SE-oriented" lines denote the EE performance of the SE maximization designs, which can be implemented via setting $\xi_u$, $\forall u,$ to be zero in problem $\cP_1$ or $\bar{\cP}_1$, as is mentioned in Sections \ref{sec:EE Optimization with Instantaneous CSIT} and \ref{sec:EE Optimization with Statistical CSI}.

In \figref{fig:CSIComparison}, the DMA weights are chosen from the complex-plane set. We compare the EE performance of the DMA-assisted uplink system versus the power budget $P_{\max}$ between the instantaneous and statistical CSI cases. As expected, the EE performance is better when the instantaneous CSI can be perfectly known in both the EE- and SE-oriented approaches. We also observe that the EE performance based on the statistical CSI is quite close to that based on the instantaneous CSI. Note that, the optimization process in the statistical CSI case is more
computationally efficient than the instantaneous CSI one. Thus, in our DMA-assisted communication scenario, the statistical CSI is a good substitute for the instantaneous CSI to maximize the system EE. In addition, \figref{fig:CSIComparison} shows that the EE performance of both EE- and SE-oriented approaches are almost identical in low and medium power regions. This is because in such regions, the circuit and the dynamic power consumption dominates. It can also be observed that the EE performance of the EE-oriented approach remains a constant while that of the SE-oriented one continues to deteriorate in the high power budget region. This phenomenon can be explained as follows. In the EE-oriented approach, there exists a saturation point of the optimal transmit power for maximizing EE. Any power that exceeds the threshold is redundant. On the contrary, the SE maximization in the SE-oriented approach always uses the full-power budget, thus resulting in the degradation of the EE performance in the high power budget region.

\subsection{EE Performance Comparison with Other Baselines}
This subsection aims to compare the EE performance between the DMA- and convectional antenna-assisted systems with fully digital and hybrid A/D architectures.
We firstly illustrate the EE models of the conventional antennas-assisted systems for both architectures. For clarity, we list the power consumption models for the considered architectures as well as the considered typical parameter setup in Tables \ref{tab:power} and \ref{tab:parameter}, respectively.
\definecolor{lightblue}{rgb}{0.93,0.95,1.0}
\begin{table}[t]
 \caption{Power Consumption for Different Architectures}\label{tab:power}
 \centering
 \footnotesize
 \ra{1.5}
\begin{tabular}{ l c c c }
  \toprule
   &   {DMAs} &   {Fully Digital} &   {Hybrid A/D}\\
  \midrule
  \rowcolor{lightblue}
  Static circuit power per users & $ W_{\text{c},u}$ & $W_{\text{c},u}$ & $W_{\text{c},u}$ \\
  Static circuit power
 at the BS  & $W_{\text{BS}}$ & $W_{\text{BS}}$ & $W_{\text{BS}}$ \\
 \rowcolor{lightblue}
 Dynamic power of RF chains & $K W_{\text{S}}$ & $M W_{\text{S}}$ & $K W_{\text{S}}$ \\

  Power of phase shifters & \ding{53}  & \ding{53}   & $K M W_{\text{p}}$ \\
  \bottomrule
\end{tabular}
\end{table}

\begin{table}[t]
 \caption{Setup of Power Parameters}\label{tab:parameter}
 \centering
 \footnotesize
 \ra{1.5}
\begin{tabular}{ l c}
  \toprule
 Parameters &  Values \\
  \midrule
  \rowcolor{lightblue}
  Static circuit power per users $ W_{\text{c},u}$ & 20 dBm \cite{2021Reconfigurable}\\
  Static circuit power at the BS $W_{\text{BS}}$  & 40 dBm \cite{2013Coordinated}\\
   \rowcolor{lightblue}
  Dynamic power consumption per RF chain $W_{\text{S}}$ & 30 dBm \cite{2013Coordinated}\\
  Power consumption per phase shifter $ W_{\text{p}}$  & 30 mW \cite{2016PhaseorSwitches}\\
  \bottomrule
\end{tabular}
\end{table}
\subsubsection{EE Model of the Fully Digital Architecture}\label{subsubsec:Con}
In the fully digital architecture-based system, each antenna element is connected with an independent RF chain \cite{2010Noncooperative,2013Hoydis,2020mmw}. For the case with instantaneous CSI, the achievable EE is given by
\begin{align}\label{eq:conven_EE}
     EE_{\text{C}} = B \frac{\log_2 \left| \bI_M + \frac{1}{\sigma^2} \sum_{u=1}^U{\bG_u \bP_u \bG_u^H} \right|}{\sum_{u=1}^{U}{( \xi_u \tr {\bP_u} + W_{\text{c},u})} + W_{\text{BS}} + M W_{\text{S}}}.
\end{align}
In \eqref{eq:conven_EE}, we use similar notations $\xi_u^{-1}$, $\tr{\bP_u}$ $W_{\text{c},u}$, $W_{\text{S}}$, and $W_{\text{BS}}$ as the EE model in \eqref{eq:power_consumption_model}. The main components of the total power consumption are listed in \tabref{tab:power}. The major difference from \eqref{eq:power_consumption_model} is that $W_{\text{S}}$ is multiplied by $M$ in \eqref{eq:conven_EE}, as the number of required RF chains is equal to that of antenna elements in the fully digital architecture. In addition, for the case with statistical CSI, a similar EE model can be obtained.
\subsubsection{EE Model of the Hybrid A/D Architecture}\label{subsubsec:A/D}
To further verify the EE advantages brought by the deployment of DMAs, we compare the DMA-assisted transmission with the fully-connected hybrid A/D architecture \cite{2019Family}. For the case with instantaneous CSI, the corresponding EE is given by
\begin{align}\label{eq:AD_EE_instantaneous}
  & EE_{\text{AD}} = \ntb
     & B \frac{ \log_2 \left| \bI_{U} + \frac{1}{\sigma^2}\sum_{u=1}^U{ \bR_n^{-1} \bW^H  \bG_u \bP_u \bG_u^H \bW} \right|}{ \sum_{u=1}^{U}{( \xi_u \tr {\bP_u} + W_{\text{c},u})} + W_{\text{BS}} + K W_{\text{S}} + KM W_{\text{p}} }.
 \end{align}
In \eqref{eq:AD_EE_instantaneous}, $\bR_n = \bW^H \bW $ where $\bW \in \C^{M \times K}$ denotes a hybrid combiner composed of an RF combiner $\bW_{\text{RF}} \in \C^{M \times K}$ and a baseband combiner $\bW_{\text{BB}} \in \C^{K \times U}$ at the BS, i.e., $\bW = \bW_{\text{RF}} \bW_{\text{BB}}$. The RF combiner $\bW_{\text{RF}}$ satisfies $|\bW_{\text{RF}}(i,j)|=1$, $i$, $j \in \{1, 2, \ldots, K\}$. In addition, $KMW_{\text{p}}$ denotes the power consumed by the phase shifters, which is the major difference of the power consumption from the DMA-assisted transmissions, as is shown in \tabref{tab:power}.  With the hybrid A/D combining circuitry, the great demand for the RF chains can be greatly reduced compared with the fully digital architecture. In addition, the statistical CSI case can be similarly modeled.

\subsubsection{EE Performance Comparison}

\begin{figure}
\centering
\subfloat[]{\centering\includegraphics[width=0.45\textwidth]{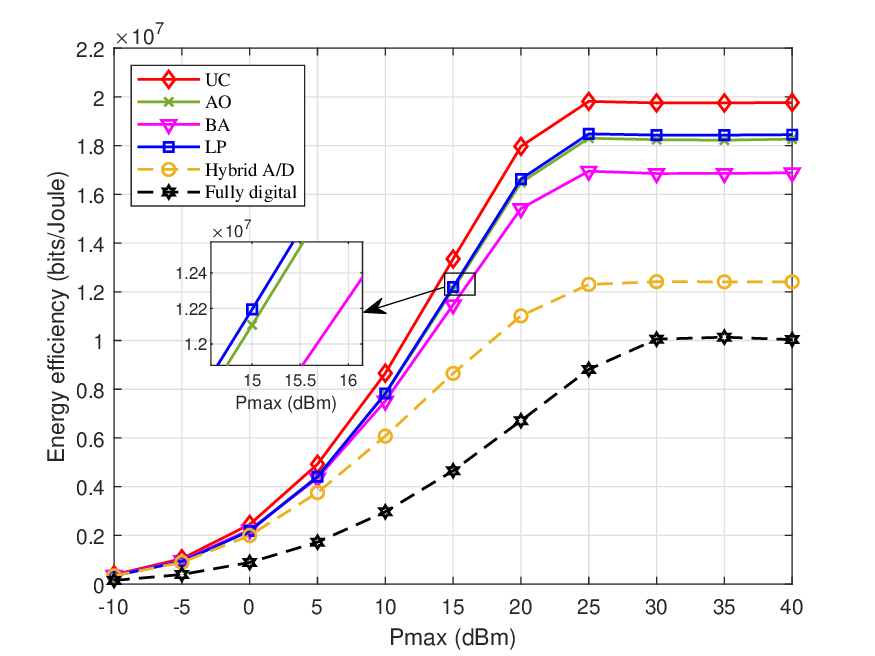}
\label{fig:EEDMA_performance}}
\hfill
\subfloat[]{\centering\includegraphics[width=0.45\textwidth]{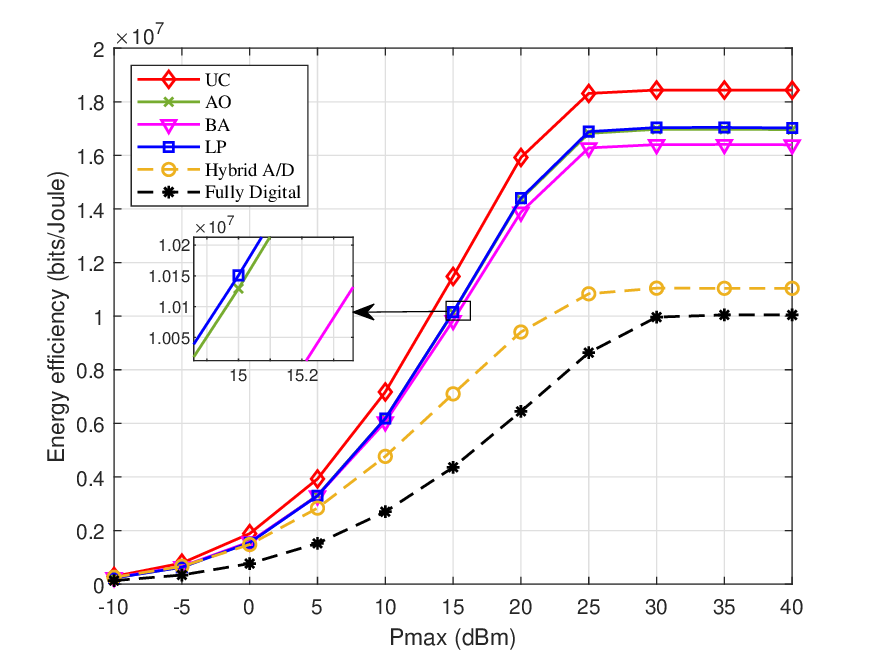}
\label{fig:EEConven_performance}}
\caption{EE performance comparison between the DMA- and conventional antennas-assisted systems: (a) instantaneous CSI; (b) statistical CSI.}
\label{fig:EEComparison}
\end{figure}

In \figref{fig:EEComparison}, we compare the EE performance between the DMA- and conventional antenna-assisted systems for both the instantaneous and statistical CSI cases. We choose the fully digital and fully-connected hybrid A/D architectures at the BS for the conventional antennas-assisted systems as the comparison baseline, whose EE models are shown above. Referring to \cite{2016PhaseorSwitches}, we assume that the power consumed by a phase shifter is $30$ mW in the hybrid A/D architecture, i.e., $W_{\text{p}} = 30$ mW.
Since the EE maximization problem of the fully digital architecture is similar to $\cP_1$ or $\bar{\cP}_1$, it can be addressed by Dinkelbach's transform. Similarly, we adopt the AO method to address the EE maximization problem with the hybrid A/D architecture. In particular, we adopt Dinkelbach's transform to optimize the transmit covariance matrices of users
and the approach proposed in \cite{2016PhaseorSwitches} to optimize the RF and baseband combiners at the BS.

From \figref{fig:EEComparison}, we can observe that the EE performance of the DMA-assisted architecture is superior to that of the conventional fully digital one, especially in the high power budget region, due to the reduced number of RF chains. In addition, the EE performance of the DMA-assisted architecture is notably better than that of the fully-connected hybrid A/D one. This is due to the fact that the hybrid A/D architecture requires additional power to support the numerous phase shifters, while DMAs do not need any additional circuitry to implement the signal processing in the analog domain. Besides, as expected, the EE performance of the hybrid A/D architecture is better than that of the fully digital architecture, which also follows from the reduced number of RF chains in the hybrid A/D architecture. In addition, we can find that the EE saturation point of the DMA-assisted architecture is shifted to the left compared with the fully digital one. This is because the DMA-assisted architecture consumes much less dynamic power with the reduction of RF chains, and then the required transmit power tends to dominate.

Comparing the four classical sets of DMA weights mentioned above, we can find that their corresponding curves scale similarly versus the transmit power budget. Among the four cases, the system EE of the complex plane case performs the best, which is attributed to the fact that the corresponding set contains the other three as subsets. We also observe that the EE performance of the continuous-valued amplitude, binary amplitude, and Lorentzian-constrained phase cases are close to the complex plane one. This phenomenon indicates that compared to the system EE in the complex plane case, the degradation of the EE performance resulting from narrowing the sets is almost negligible. It shows the possibility of a simpler implementation to achieve the comparable channel capacity and EE performance with the continuous-valued amplitude, binary amplitude, and Lorentzian-constrained phase sets. In fact, implementing the binary amplitude weight-based DMAs is much simpler, making it a more appealing solution among the four kinds for future studies.

\subsection{Effect of the Number of Microstrips}
\begin{figure}
\centering
\includegraphics[width=0.45\textwidth]{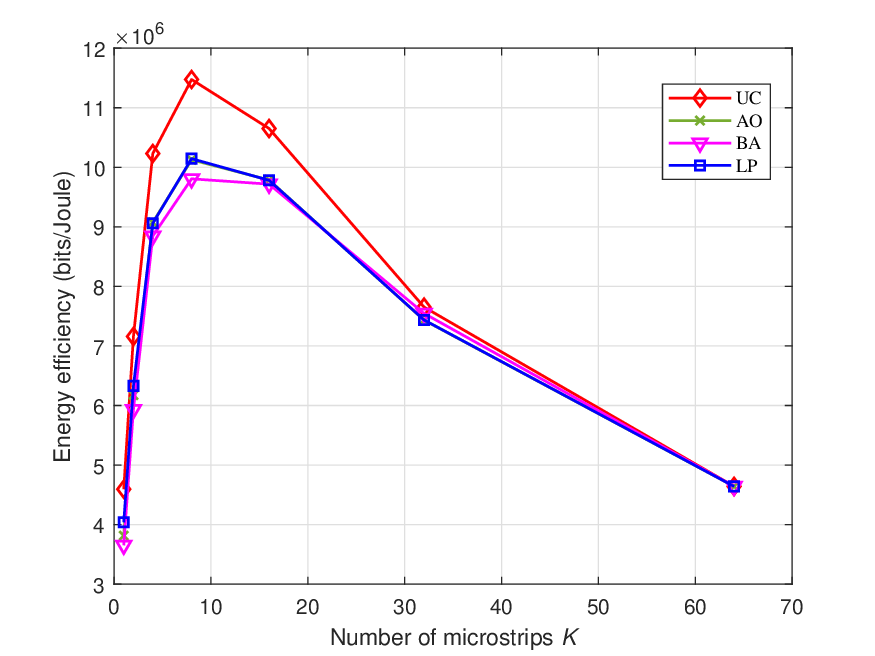}
\caption{EE performance comparison versus the number of microstrips $K$ with a fixed total number of metamaterial elements.}
\label{fig:MicrostripsComparison}
\end{figure}
In \figref{fig:MicrostripsComparison}, we evaluate the effect of the number of microstrips on the EE performance of the DMA-assisted communications. We fix the number of metamaterial elements as $M = 64$, set the transmit power consumption budget $P_{\max}$ as 15 dBm, and evaluate the EE performance of the DMA-assisted system for $K \in [1,64]$. As is shown in \figref{fig:CSIComparison}, the EE performance based on the statistical CSI is close to that based on the instantaneous CSI. Thus we focus on the EE performance based on the statistical CSI here.

From \figref{fig:MicrostripsComparison}, we note again that the achievable EE performance based on the continuous-valued amplitude, binary amplitude, and Lorentzian-constrained phase cases are close to each other and closely follow the complex plane one. We can also observe that, as the number of microstrips increases, the EE performance firstly rises to a peak and then decreases. This phenomenon is related to two main factors. Firstly, since the system SE performance mainly depends on the number of RF chains, the system SE will be improved as the number of RF chains increases. Secondly, the dynamic power consumption of RF chains will increase as the number of RF chains increases. Note that the number of RF chains is equal to that of microstrips in the DMA-assisted architecture. Then, for small $K$, the first factor dominates the EE performance, i.e., the system SE increases as the number of microstrips increases, thus resulting in the improvement of the system EE. On the contrary, for large $K$, the second factor dominates the EE performance. Specifically, the dynamic power consumption of RF chains, which is proportional to the number of microstrips, dominates for large $K$. Therefore, the EE performance decreases as the number of microstrips increases. This observation implies that in practical implementation, we need to select the number of microstrips to strike a balance between the power consumption and SE gain to improve the EE performance in the DMA-assisted communications.

\subsection{Impact of Imperfect CSI}

\begin{figure}
\centering
 \subfloat[]{\centering\includegraphics[width=0.45\textwidth]{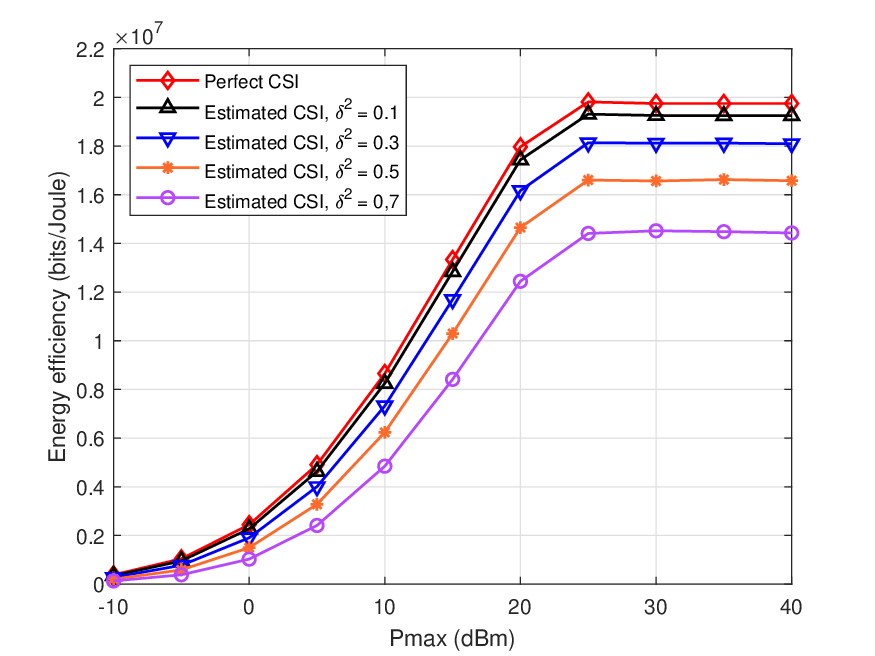}} \label{fig:ICSI_error} \hfill
 \subfloat[]{\centering\includegraphics[width=0.45\textwidth]{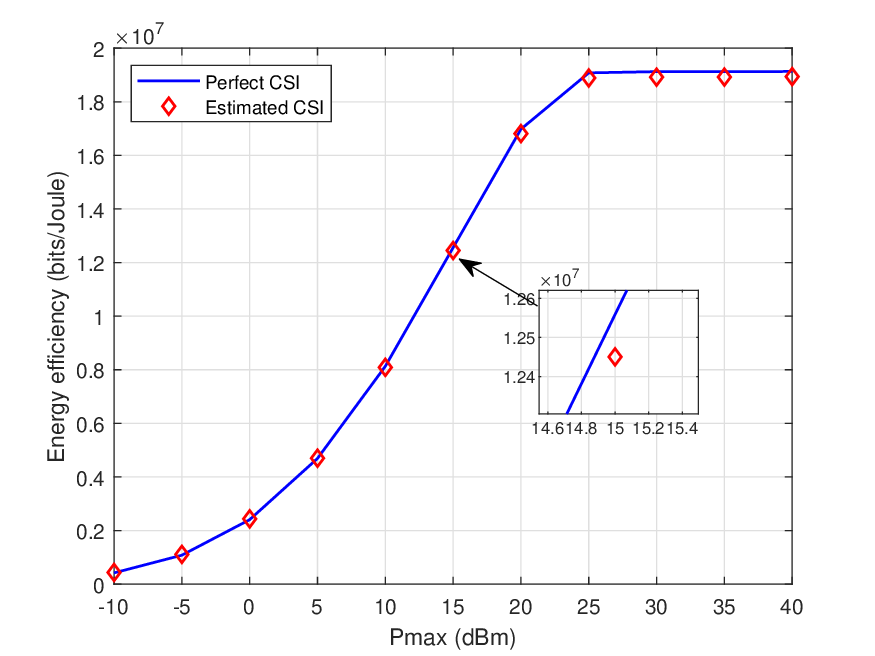}} \label{fig:SCSI_error}
\caption{EE performance of the proposed approach with perfect and imperfect CSI: (a) instantaneous CSI; (b) statistical CSI.}
\label{fig:error}
\end{figure}

In this subsection, we evaluate the impact of imperfect CSI on the performance of the proposed algorithms. We firstly consider the instantaneous CSI case. In particular, we adopt the imperfect instantaneous CSI model given by \cite{2020WangChannel}
\begin{align}\label{eq:imisntcsi}
    \bar{\bG}_u  = \bG_u + \bE_u,
\end{align}
where $\bar{\bG}_u \in \C^{M \times N_u}$ denotes the imperfectly obtained CSI of user $u$, and $\bE_u$ is the CSI error matrix with  complex-valued Gaussian entries i.i.d. as $\cC\cN(0, \delta_u^2)$, where $\delta_u^2$ describes the inaccuracy of obtained CSI. Assuming that $\delta_u^2=\delta^2$ for clarity, we compare the EE performance of the proposed algorithm with different CSI uncertainty $\delta^2$ in \figref{fig:error}(a). We can observe that for the instantaneous CSI case, the performance decreases as $\delta^2$ increases, and thus the robust transmission design for DMA-assisted systems with the CSI uncertainty taken into account will be of practical interest.

For the statistical CSI case, we assume that the statistical CSI is estimated via averaging over 50 instantaneous channel realizations via e.g., channel sounding. The EE performance of the proposed approaches with the exact and the estimated statistical CSI are presented in \figref{fig:error}(b). It can be observed that the performance loss using the estimated statistical CSI is almost negligible. This phenomenon indicates the robustness of the statistical CSI-based approach \cite{2009Statistical}.

\section{Conclusion}\label{sec:conclusion}

In this paper, we studied the EE performance optimization of the DMA-assisted massive MIMO uplink communications, considering both the cases of exploiting the instantaneous and statistical CSI. Specifically, we developed a well-structured and low-complexity framework for the transmit covariance design of each user and the DMA configuration strategy at the BS, including the AO and DE methods, as well as Dinkelbach's transform. Based on our algorithm, the DMA-assisted communications achieved much higher EE performance gains compared to the conventional large-scale antenna array-assisted ones, especially in the high power budget  region. The results also showed that the EE performance based on DMAs could be further improved by adjusting the number of microstrips. In the future work, robust DMA-assisted transmission design incorporating the imperfect CSI effect will be of practical interest.

\end{document}